\documentclass[twocolumn,secnumarabic,amssymb, nobibnotes, prb]{revtex4-1}
\pdfoutput=1
\usepackage{color}
\usepackage{graphicx}
\usepackage{multirow}
\usepackage{subfigure}

\newcommand{\C}[1]{{}} 

\begin{document}
\title{First-principles investigation of incipient ferroelectric trends of rutile TiO$_2$ in bulk and at the (110) surface}%
\author{A. Gr\"unebohm}%
\email{anna@thp.uni-due.de}
\author{P. Entel}
\affiliation{Faculty of Physics and CENIDE, University of Duisburg-Essen,  47048 Duisburg, Germany}
\author{C. Ederer} 
\affiliation{Materials Theory, ETH Z\"urich, Wolfgang-Pauli-Str.\,27, 8093 Z\"urich, Switzerland}
\altaffiliation[Previous address: ]{School of Physics, Trinity College, Dublin 2, Ireland}
\date{May, 2012}%
\begin{abstract}
The ferroelectric trends of rutile (TiO$_2$) in bulk and at the (110)
surface are investigated by means of {\it ab initio} density
functional theory.  We discuss the underlying mechanism of the
incipient ferroelectric behavior of rutile in terms of Born effective
charges, which we decompose in individual contributions by means of
maximally localized Wannier functions. We show that a ferroelectric
phase can be stabilized for a variety of different lattice distortions
which all enlarge the shortest Ti-O distance, even if the longer
apical Ti-O bond is simultaneously shortened. At the (110) surface,
the ferroelectric trends are modified compared to the bulk, but
nevertheless ferroelectric phases with large polarization even in the
topmost surface layer can be stabilized by uniaxial strain.
 \end{abstract}
 \pacs{77.80.bn,77.84.Bw,77.55.fp}
 \maketitle

\section{Introduction}
TiO$_2$ in the rutile structure has been classified as an incipient
ferroelectric material, based on the experimental observation that the
dielectric permittivity is exceptionally large and increases with
decreasing temperature, although no ferroelectric transition occurs
even for $T=0$~K.\cite{Traylor,Samara,Parker}
The realization of ferroelectric TiO$_2$, or even the possibility to
efficiently tune its dielectric constant, would be of great
technological importance, since TiO$_2$ is a cheap material with well
optimized processing methods, which is used for a variety of
applications such as optical coating,\cite{Siefering} solar
cells,\cite{Graetzel} or sensors.\cite{Gas} Many of these applications
are based on properties involving the rutile (110)-surface, which is
the technologically most relevant and therefore best-studied rutile
surface.\cite{Bredow, Bates, Diebold2,Murugan,Ramamoorthy}

Previous theoretical {\it ab initio} studies have shown that in the
corresponding bulk material the anomalous behavior of the dielectric
constant is a consequence of a low-frequency polar (A$_{2u}$) phonon
mode,\cite{Lee} which softens under negative pressure or uniaxial
tensile strain along the (001)-axis.\cite{Montanari2, Montanari,Hong}
Liu {\it et al.}  classified the phase transition between the
paraelectric and ferroelectric phases to be of second order.
Since the resulting magnitude of the ionic off-centering as well as
the electric polarization are in the same range as for prototypical
ferroelectrics such as PbTiO$_3$ or BaTiO$_3$, an enforced
ferroelectric transition would lead to a technologically relevant
ferroelectric material. Furthermore, Mitev {\it et al.} found a large
region within the Brillouin zone with a low lying phonon branch which
may be linked to the structural relaxation of the rutile (110)
surface.\cite{mitev} These authors also pointed out a softening of
this mode under uniaxial strain in both [001] and [110] directions,
and found that the ferroelectric A$_{2u}$-mode is quite insensitive
towards [110]-oriented strain. In contrast, we have shown more
recently that the A$_{2u}$-mode indeed also softens under [110]
strain, but that a different polar mode, an E$_u$ mode polarized along
[110], becomes even more unstable than the A$_{2u}$ mode under this
strain orientation.\cite{110} Under uniaxial tensile [110] strain, we
have thus found a second order phase transition to a ferroelectric
phase with polarization along [110].\cite{Bordeaux} In agreement with
these results, a softening both of the A$_{2u}$ mode and of the lowest
E$_u$ mode has also been found in Ref.~\onlinecite{Hong} for tensile
uniaxial [001] strain and for an isotropic lattice expansion within
the (100)/(010) plane. Since the strain dependence of the two polar
modes is different, this opens up the interesting possibility to
``strain engineer'' the polarization direction.

It is well known that ferroelectric or dielectric properties can be
significantly modified at surfaces in comparison with the
corresponding bulk material. First of all, a polarization along the
surface normal induces a depolarizing field, and is therefore often
suppressed in thin films (see, e.g., Ref.~\onlinecite{Rabe}). Second,
the reduced coordination of the surface atoms may prevent or enhance
ferroelectric trends,\cite{Ravikumar,Padilla,Meyer2} as the dipolar
interaction and the short-range repulsion are both modified.  Finally,
the surface-induced atomic relaxations can also modify the bonding and
thus the ferroelectric properties. For example, it has been discussed
whether the surface relaxation and the unsaturated bonds at the
surface are able to destabilize the paraelectric state, leading to a
finite electric polarization in the related incipient ferroelectric
material SrTiO$_3$.\cite{Ravikumar,Padilla,Bickel} We also note that a
ferroelectric transition has been observed experimentally in strained
SrTiO$_3$ films.\cite{Haeni} Therefore, an obvious approach to achieve
a ferroelectric transition in rutile is the growth of strained thin
TiO$_2$ films.

In view of this, it is essential to test the modifications of the
ferroelectric properties of the strained TiO$_2$ (110) surface, as
considerable differences may appear compared to the bulk
case. Nevertheless, to the best of our knowledge no such investigation
of the ferroelectric trends of this TiO$_2$ surface has been
undertaken until now.

In the present paper we first present a systematic comparison of the
effects of different lattice modifications on the ferroelectric
properties of bulk rutile TiO$_2$. We show that each lattice
modification which affects the nearest neighbor Ti-O distances within
the crystal, can stabilize a ferroelectric phase polarized along
(001). Additionally, we discuss the fundamental mechanisms for the
incipient ferroelectric behavior of rutile TiO$_2$ in the framework of
Born effective charges. We then present the {\it{ab initio}}
investigation of the ferroelectric trends of the TiO$_2$ rutile
surface. Although we find strong surface-induced relaxations and
modifications of the electronic structure, which have a considerable
influence on the ferroelectric properties, the strong ferroelectric
tendencies of the bulk material are essentially retained at the
surface. In addition, two different ferroelectric states with
polarization within the surface plane can be stabilized for strained
TiO$_2$ films. Here, local dipoles both along (001) and along
($\bar{1}10)$ can be stabilized within the surface at 2~\% uniaxial
strain. Furthermore the magnitude of the polarization increases as the
strain is enlarged to 5~\%.

This paper is organized as follows. In Sec. \ref{sec:comp} we present
our computational methods and all related technical details. In
Sec.~\ref{sec:bulk} we discuss the ferroelectric trends of bulk rutile
in some detail. This is essential in order to make reliable
predictions on the ferroelectric trends of the surface discussed in
the later sections. For this purpose, the ferroelectric trends under
different lattice modifications and a discussion of the underlying
mechanisms will be presented in Sec.~\ref{subsec:modes} and
Sec.~\ref{subsec:born}, respectively. We outline the ferroelectric
trends of the rutile (110) surface in Sec.~\ref{sec:surface}, both in
free films and for films which are clamped to an idealized substrate
(Sec.~\ref{subsec:unstrained}).  Finally, we show that different
ferroelectric distortions can be stabilized in the surface under
tensile uniaxial strain and discuss the resulting displacement
patterns in Sec.~\ref{subsec:strained}. Conclusions and outlook on
future work are summarized in Sec.~\ref{sec:outlook}.

\section{Computational details}
\label{sec:comp}

The electronic structure of TiO$_2$ is calculated using
first-principles density functional theory.\cite{kohnsham} Most
calculations are performed by employing the Vienna Ab Initio
Simulation Package (VASP 5.2.2)\cite{Kresse1} based on projector
augmented wave pseudo-potentials.\cite{Blochl} Born effective charges
and the dielectric permittivity are calculated within VASP using
density functional perturbation theory.\cite{bornstoer}
The generalized gradient approximation (GGA) in the formulation of
Perdew, Burke and Ernzerhof (PBE)\cite{PBE} is used for the exchange
correlation energy, with $4s3d$ ($2s2p$) electrons treated as valence
for Ti (O). Although small inaccuracies may result from the use of
such a soft Ti pseudopotential,\cite{Muscat2} we could not find
qualitative differences in the description of the electronic and
dielectric properties compared to an enlarged valence basis set and
thus stick to the minimal valence electron basis.
A planewave cutoff of 500~eV, an energy convergence better than
$10^{-7}$\,eV, and at least $6\times6\times6$ ($5\times13\times3$)
k-points constructed with the Monkhorst-Pack\cite{Monkhorst} scheme
for bulk (films) guarantee sufficient accuracy of our calculations.
A Gaussian smearing of 0.1~eV is used for most calculations in order
to improve the convergence with respect to the number of
k-points. This smearing is reduced to 0.05~eV for the determination of
the bulk Born effective charges, and the linear tetrahedron method is
used for the determination of the harmonic frequencies.
Atomic positions are optimized until the residual forces are smaller
than 0.001\,eV/{\AA} (0.005\,eV/{\AA}) for bulk (surface)
calculations. For the surface calculations at least 10~{\AA} of vacuum
is used between the periodic images of the films.

The Plane-Wave Self-Consistent Field (PWscf) code, which is part of
the QuantumEspresso package,\cite{espresso} in combination with
Wannier90,\cite{Wannier90} is used to construct maximally localized
Wannier functions (MLWFs)\cite{Wannier1} for the orbital decomposition
of the Born effective charges of bulk rutile TiO$_2$. For this
purpose, a $\Gamma$-centered equidistant $14\times14\times 20$
$k$-point grid, a plane-wave cutoff of 35~Ry ($\approx$ 476~eV) for
the wave-functions (420~Ry for the charge density), and an energy convergence of $10^{-8}$~Ry ensure
sufficient accuracy. The Ti $3s$ and $3p$ semi-core states are treated
as valence electrons in the PWscf calculations.  For the MLWF
decomposition of the surface, the $k$-point grid is reduced to
$6\times12\times 2$.  MLWFs are constructed separately for valence and
semi-core states, based on appropriate initial projections, and the
total spread is always converged better than $10^{-10}$\,\AA$^2$
($10^{-7}$\,\AA$^2$) per Wannier function for the bulk (surface).

\section{Bulk rutile}
\label{sec:bulk}
In order to provide a well-defined reference for the interpretation of
our results for the TiO$_2$ rutile (110) surface in
Sec.~\ref{sec:surface}, we first discuss results for the corresponding
bulk system.

\subsection{Structural properties}
\label{subsec:struc}
\begin{figure}
\includegraphics[width=0.25\textwidth]{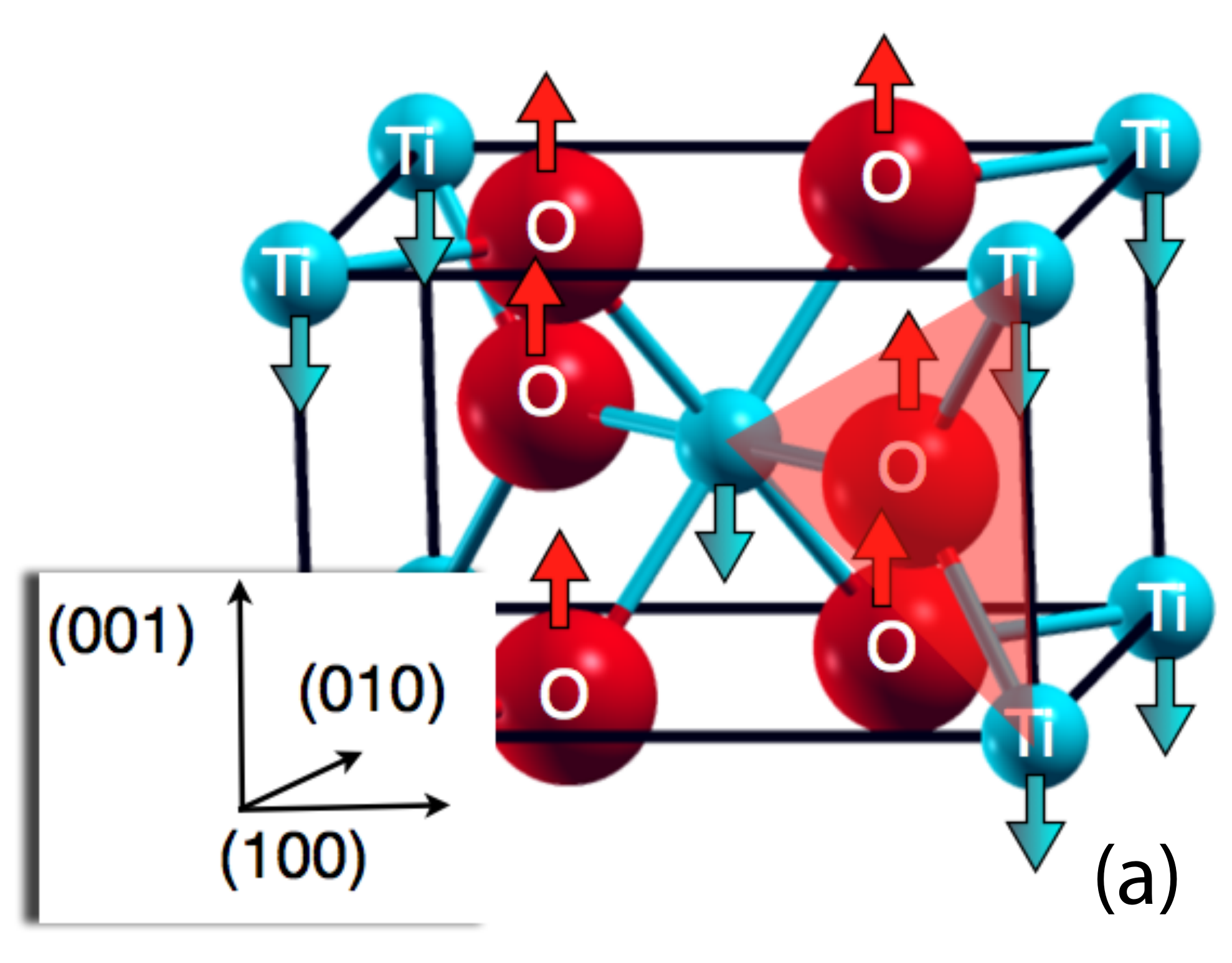}
\includegraphics[width=0.2\textwidth]{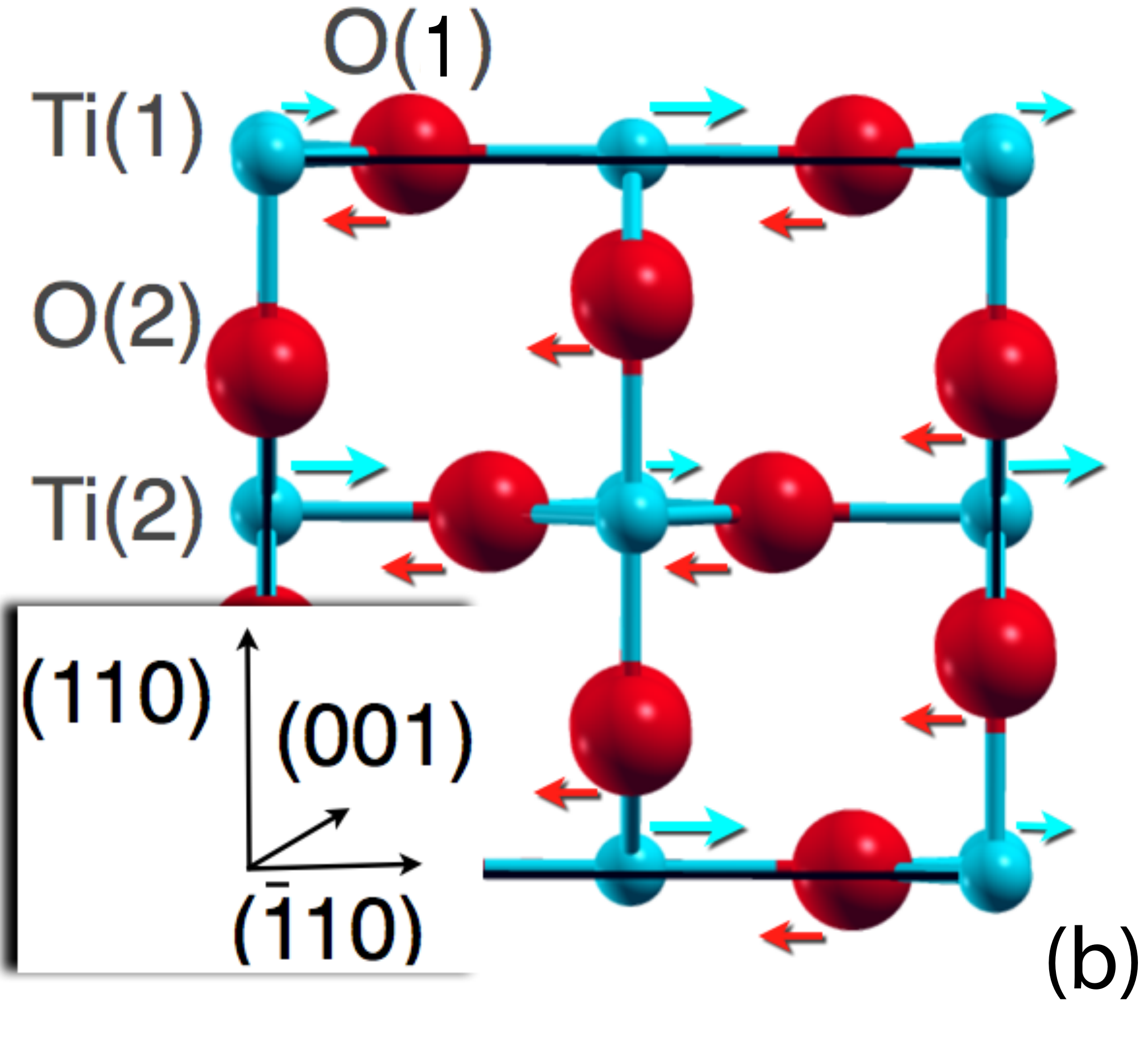}
\caption{(Color online) Atomic structure of rutile. Ti (blue, light),
  O (red, dark).  (a) Primitive unit cell. Arrows mark the atomic
  displacement of the A$_{2u}$ phonon mode. The atomic structure can
  be decomposed into O-Ti$_3$ units, see red triangle.  (b)
  $\sqrt{2}\times1\times\sqrt{2}$ cell in [$\bar{1}$10], [001], [110]
  coordinates. Ti(1)/Ti(2) /O(1)/O(2) mark symmetry-inequivalent
  atomic classes under $[\bar{1}10]$ strain. Arrows mark the polar
  atomic displacements under $[\bar{1}10]$ strain, see
  Ref.~\onlinecite{Bordeaux}.}
\label{fig:rutil}
\end{figure}

\begin{table}
\caption{Bulk lattice parameters and Born charges $Z^*$ (see
  Sec.~\ref{subsec:born} for further details) of rutile TiO$_2$.
  Results obtained in the present work (top rows) are compared to
  literature data (bottom rows).}
\begin{tabular}{lcccccccc}
\hline
\hline
&$a$ [{\AA}]&$c/a$&$u$&Z$^*_{||}$&Z$^*_{\perp}$&Z$^*_{(001)}$\\
\hline
GGA\footnote{VASP}&4.664&0.637&0.305&7.60&5.36&8.15\\
GGA\footnote{PWscf; structural parameters fitted to VASP results.}& & & &7.25&-&7.73\\
\hline
 LDA \cite{mitev}&4.572&0.644&0.304&7.49&5.43&7.77\\
 LDA\cite{Shojaee}&4.535&0.641&0.303&7.27&5.25&7.32\\
 GGA\cite{Shojaee}&4.637&0.638&0.305&7.22&5.18&7.52\\
exp.\footnote{4.2 K, Ref.~\onlinecite{Samara}} &4.587&0.644&0.305&-&-&-\\
\hline
\hline
\end{tabular}
\label{tab:const}
\end{table}

The atomic structure of TiO$_2$ in its rutile morphology is tetragonal
with $P4_2mnm$ symmetry, see Fig.~\ref{fig:rutil}. Besides the lattice
constant $a$ and the tetragonal $c/a$-ratio, an internal parameter $u$
defines the Ti-O distances.  Each Ti-atom is 6-fold coordinated by
O-atoms with 4 short equatorial Ti-O bonds (Ti-O$_{eq}$),
$d_{eq}=\left|(2u-1,2u-1,c/a)\right|\cdot a/2$, and two long apical
Ti-O bonds (Ti-O$_{ap}$), $d_{ap}=\left|(u,u,0)\right|\cdot a$.

It has been shown that the ferroelectric trends of bulk rutile depend
critically on the lattice
parameters.\cite{Liu,Montanari,Bordeaux,110,mitev} Because of this, a
correct description of the atomic structure is essential. Our
calculated lattice parameters, see Table~\ref{tab:const}, and the
calculated pressure dependence of these quantities, see
Tab.~\ref{tab:p}, fit better to experimental results than calculations
based on the commonly used local density approximation
(LDA). Nevertheless, the LDA error is overcorrected, leading to an
overestimation of the lattice constant by 1.6\,\% compared to low
temperature experimental results, which may also lead to a
quantitative overestimation of the ferroelectric trends. We note that,
while an artificial destabilization of the paraelectric phase for GGA
potentials has been found in Ref.~\onlinecite{Montanari2}, we could
not reproduce this in our calculations, in agreement with
Ref.~\onlinecite{Shojaee}, see Tab.~\ref{tab:fit}. These differences
can most likely be attributed to the different pseudopotentials and
technical details such as the used basis functions, see also the
related discussion for the incipient ferroelectric material
SrTiO$_3$.\cite{Evarestov}

\begin{table}
\caption{Modification of the lattice parameters under external
  pressure $P$. Our values obtained with GGA-potentials are opposed to
  values obtained with LDA potentials from
  Ref.~\onlinecite{Montanari}, and experimental data is given for
  comparison. For each quantity the calculated pressure dependence has
  been fitted by $f(P)=b_0+b\cdot P$. The values for zero pressure,
  $b_0$ (left), and the slopes $b$, in units 10$^{3}$~GPa$^{-1}$ for
  $u$ and $c/a$, in units of 10$^{3}$~GPa$^{-1}$\AA\ otherwise (right)
  are listed.}
\label{tab:p}
\begin{tabular}{rccccccc}
\hline
\hline
& \multicolumn{3}{c}{b$_0$ }&~~~~& \multicolumn{3}{c}{b}\\
&LDA\cite{Montanari}&Here&Exp.\cite{Montanari}&&LDA\cite{Montanari}&Here&Exp.\cite{Montanari}\\
\hline
a [\AA]&4.55&4.66&4.59&&-7.73&-9.1&-8.62\\
c [\AA]&2.92&2.97&2.96&&-2.41&-2.2&-2.14\\
u~~~~~~&~~0.304&~~0.305&-&&-0.14&-0.2&-\\
c/a~~~~~&0.64&0.64&0.64&&~0.56&~0.8&~0.79\\
TiO$_{eq}$[\AA]&1.93&1.97&-&&-1.70&-1.5&-\\
TiO$_{ap}$[\AA]&1.96&2.01&-&&-4.27&-5.4&-\\
\hline
\hline
\end{tabular}
\end{table}

\subsection{Polar phonon modes} 
\label{subsec:modes}
\begin{figure}
\includegraphics[width=0.5\textwidth]{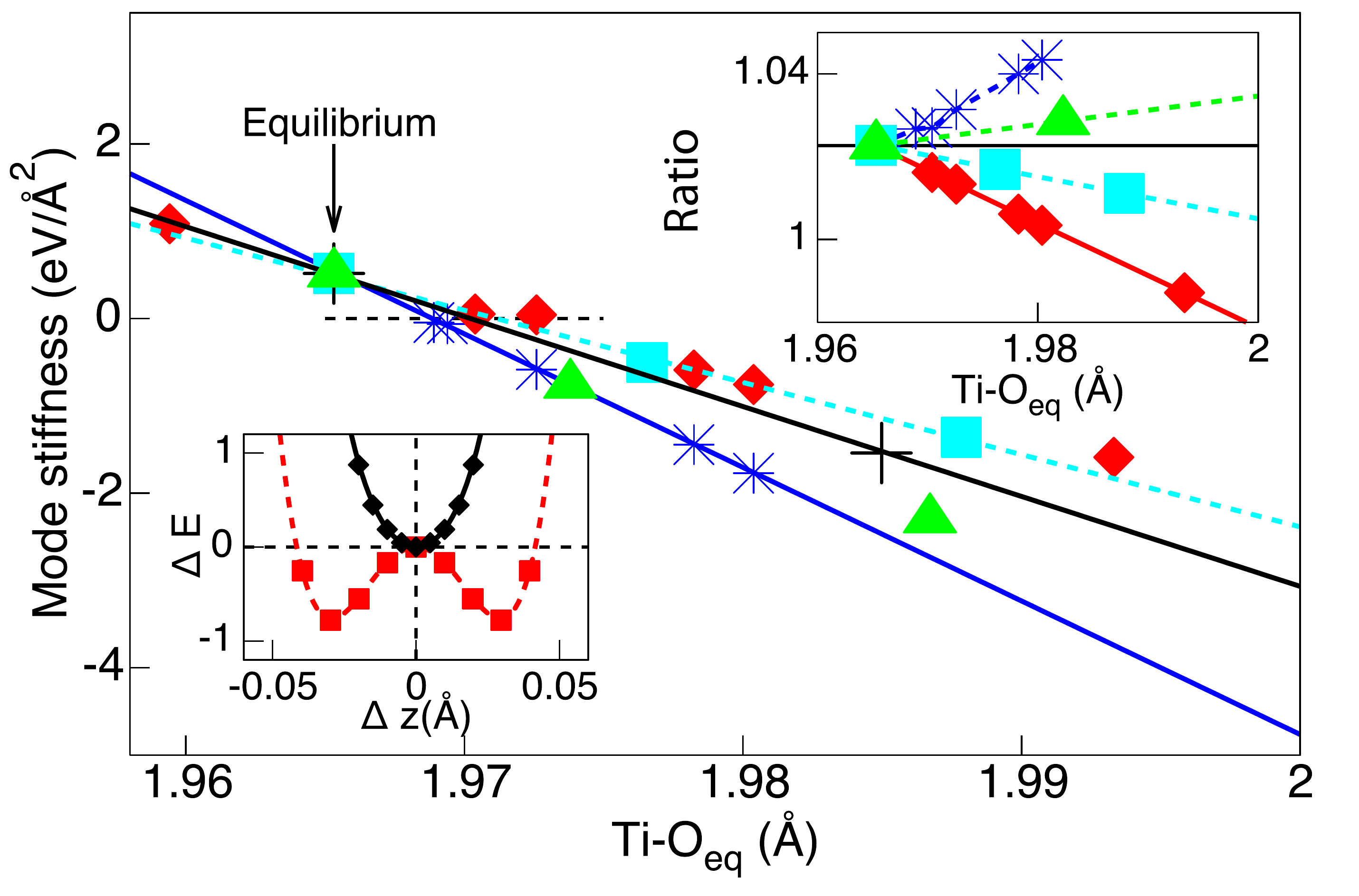}
\caption{(Color online) Stiffness $s$ of the A$_{2u}$ mode for
  different modifications of the equilibrium crystal structure:
  External pressure (blue, stars); $u$-variation (red, diamonds);
  [001] strain (cyan, squares); [$\bar{1}$10] strain (green,
  triangles); isotropic lattice expansion (black, crosses). Left
  inset: Energy difference (meV/2 f.u.) for a static atomic
  displacement at the equilibrium lattice constants (black, diamonds)
  and under 1~\% tensile [001] strain (red, squares). $\Delta z$ is
  the relative Ti-O displacement in [001] direction. Right inset:
  Ratio of the Ti-O$_{ap}$ and Ti-O$_{eq}$ distances under the imposed
  lattice modifications.}
\label{fig:ezbulk}
\end{figure}

In the following, the relative stability between the paraelectric
state and a ferroelectric state with polarization along [001] will be
discussed in the framework of the frozen phonon approach, {\it cf.}
Refs.~\onlinecite{Montanari,Montanari2,Liu,mitev}. For this purpose
the atoms are successively displaced along the eigenvector $\vec{Y}$
of the A$_{2\text{u}}$ phonon mode and the total energy is monitored.

The eigenvector of the A$_{2u}$ mode is fully determined by symmetry
as it is the only polar mode along [001],\cite{Lee} and a gradual
shift of amplitude $A$ along the eigenvector $\vec{Y}$ is thus given
by the displacements $A\cdot (0,0,1)$ for Ti and $A\cdot
(0,0,-m_{Ti}/2m_{O})$ for O-atoms,
respectively. Here, $m_i$ represent the Ti and O masses. 

While the paraelectric state is stable at the equilibrium lattice
constants, the energy surface flattens around $A=0$ for tensile [001]
strain, or under an isotropic volume expansion. In both cases the
A$_{2\text{u}}$ mode softens proportional to the imposed lattice
modification, and two ferroelectric minima appear at a critical value
of the lattice modification,\cite{Montanari,Liu,mitev} see left inset
in Fig.~\ref{fig:ezbulk}. Although so far no such transition has been
observed experimentally, a considerable modification of the $A_{2u}$
mode frequency under external pressure has been measured.\cite{Samara}

In order to discuss these ferroelectric trends in more detail, we
expand the energy change per primitive cell during the polar
displacement, $\Delta E(A)$, up to 4$^{th}$ order with respect to
$|A \cdot \vec{Y}|$,
\begin{equation}
\label{eq:4}
\Delta E(A)=\frac{s}{2}A^2\left|\vec{Y}\right|^2+\kappa
A^4\left|\vec{Y}\right|^4 \,.
\end{equation}
For a qualitative discussion we use the stiffness $s$, which
determines the curvature of the energy at the paraelectric
state. 

\begin{table}
\caption{Coefficients of total energy fits, Eq.~\ref{eq:4}, for atomic
  displacements along the A$_{2u}$ mode and harmonic frequencies
  $\omega_0$ in comparison with literature data.  Top part: external
  pressure; Center: tensile [001] strain; Bottom part: tensile
  [$\bar{1}$10] strain.}
\begin{tabular}{c c c c c c c c c}
\hline
\hline
&\multicolumn{2}{c}{Equation \ref{eq:4}}&\multicolumn{4}{c}{Harmonic frequencies}\\
&$s$&$\kappa$&PBE& LDA \cite{Montanari}&PBE~\cite{Montanari2}&LDA~\cite{Liu}\\
&(eV/{\AA}$^2$)&(eV/{\AA}$^3$)&\multicolumn{4}{c}{$\omega_0$(cm$^{-1}$)}\\
\hline
 Eq.&~~0.51&21.89&~~91&150~&$i$\,86&170\\
 -2 GPa&~~0.14&21.36&~~47&100&-&150\\
 -5 GPa&~-0.58&21.71&$i$~97&-&-&-\\
\hline
1~\%&~-0.50&20.74&~$i$\,90&130&-&-\\
2~\%&~-1.36&17.85&$i$\,148&100&-&-\\
5~\%&-4.00&15.29&$i$\,254&$i$\,125~~&-&-\\
\hline
2~\%&-0.76&-&$i$\,110&-&-&-\\
5~\%&-2.30&&$i$\,192&-&-\\
\hline
\hline
\end{tabular}
\label{tab:fit}
 \end{table}

The stiffness is positive for a stable paraelectric state and changes
its sign as the ferroelectric state becomes the energetic ground
state. Figure~\ref{fig:ezbulk} depicts the variation of the stiffness
as a function of the Ti-O$_{eq}$ bond length for the following
modifications of the equilibrium crystal structure: ({\it i)} external
pressure, ({\it ii)} isotropic lattice expansion, ({\it iii)} [001]
strain, ({\it i}v) [$\bar{1}10$] strain, and (v) variation of the
ratio between TiO$_{eq}$ and TiO$_{ap}$ bond lengths at constant
volume by a variation of the internal parameter $u$. Here, for ({\it
  i)}, all lattice parameters have been optimized at a given pressure,
whereas neither the lattice vectors nor the internal parameter $u$
have been optimized for all other cases.
  
It can be noticed that we obtain a stable ferroelectric state
polarized along [001] for each tested lattice modification which
enlarges the Ti-O$_{eq}$ bond, see Fig.~\ref{fig:ezbulk}. Even an
artificial modification of the internal parameter $u$, which enlarges
the Ti-O$_{eq}$ bonds and reduces the Ti-O$_{ap}$ distances without
lattice expansion, stabilizes the ferroelectric phase.

For the case of [001] strain, it has been discussed in
Ref.~\onlinecite{Montanari}, that the large short-range repulsion,
which appears under a relative displacement of the two sublattices,
prevents a ferroelectric transition for the equilibrium structure, but
that this contribution to the total energy decreases with increasing
strain.  Our systematic comparison of different lattice modifications
further shows that it is sufficient to increase the Ti-O$_{eq}$
distance in order to stabilize the ferroelectric phase.  Most likely,
the relative displacements of these ions contribute most to the
short-range repulsion as they correspond to the shortest Ti-O
distance.

Even though all the various lattice modifications destabilize the
paraelectric phase, the slopes of $s$ as a function of the Ti-O$_{eq}$
bond length differ considerably for the different cases.  Here, the
differences in the Ti-O$_{ap}$ distances result in different
contributions of these atomic pairs to the short-range repulsion.
Additionally, the dipolar interaction is reduced as the ratio of the
nearest and next-nearest Ti-O distances changes, see
Sec.~\ref{subsec:born}.  As the dipole-dipole interaction is the
second large contribution to the total energy, that is affected by
ferroelectric
displacements,\cite{Cochran:1960,Ghosez/Gonze/Michenaud:1996} its
reduction reduces the ferroelectric trends considerably.  Because of
this, the slopes of $s$ are correlated with the ratio of the two
classes of Ti-O bond lengths, see right inset of
Fig.~\ref{fig:ezbulk}.  On the one hand, the configuration for which
the atomic structure has been optimized under external pressure, shows
the steepest $s$ in Fig.~\ref{fig:ezbulk}, as an increase of the ratio
of the TiO$_{ap}$ and TiO$_{eq}$ distances is superimposed to the
lattice expansion, see Tab.~\ref{tab:p}.  On the other hand, $s$ is
notably flatter if the ratio of the two Ti-O distances decreases,
e.g., under [001] strain or especially for a modification of $u$,
which induces the largest decrease of the bond length ratio.

For a quantitative comparison of the obtained mode stiffness, the
harmonic frequencies $\omega_{0}$ can be obtained from Eq.~\ref{eq:4},
\begin{equation}
\Delta E=\frac{\omega_{0}^2}{2}A^2\left(2m_{Ti}+4\cdot
m_{O}\cdot\left(\frac{m_{Ti}}{2\cdot m_O}\right)^2\right)\,.
\end{equation}
In this way, we obtain a frequency of the A$_{2\text{u}}$ mode at the
equilibrium structure, see Tab.~\ref{tab:fit}, which is within the
spread of different theoretical estimates and below the experimental
value of 173~cm$^{-1}$.\cite{Traylor} See also the discussion in
Sec.~\ref{subsec:struc}. Apart from these quantitative differences,
similar qualitative trends for the mode softening under a lattice
modification (pressure and [001] strain) are obtained within the
different theoretical studies.

In summary, a ferroelectric phase polarized along [001] can be
stabilized for each tested expansion of the Ti-O$_{eq}$ bond. We will
see in Sec.~\ref{subsec:strained} that this result can qualitatively
be transferred to the ferroelectric trends of the surface.

\begin{table}
\caption{Modification of the internal parameter $u$, polar
  displacements $z$ of the Ti and O ions in {\AA}, and the energy
  difference $\Delta E$ relative to the optimized paraelectric state
  in meV/atom for different imposed tensile strains. Ti(1)/O(1)
  corresponds to Ti-O$_{ap}$ bonds along [110], Ti(2)/O(2) to
  Ti-O$_{ap}$ bonds along [$\bar{1}10]$, see
  Fig.~\ref{fig:rutil}(b). Upper part: Ferroelectric state polarized
  along [001]; Lower part: Ferroelectric state polarized along
  [$\bar{1}$10].}
\begin{tabular}{ccccccccc}
\hline
\hline
Pol.&Strain&$u_{[\bar{1}10]}$&$u_{[110]}$&$z_{Ti(1)}$&$z_{Ti(2)}$&$z_{O(1)}$&$z_{O(2)}$&~$\Delta$E\\
\hline
\multirow{4}{*}{[001]}&$[001]_{+2\%}$&\multicolumn{2}{c}{0.305}&\multicolumn{2}{c}{0.05}&\multicolumn{2}{c}{-0.05}&~-1~~\\
&$[001]_{+5\%}$&\multicolumn{2}{c}{0.303}&\multicolumn{2}{c}{0.08}&\multicolumn{2}{c}{-0.13}&-11~~\\
&[$\bar{1}10]_{+2\%}$&\multicolumn{2}{c}{0.305}&0.08&0.01&\multicolumn{2}{c}{-0.02}&~-1~~\\
&[$\bar{1}10]_{+5\%}$&0.035&0.036&0.14&0.02&-0.04&-0.03&~-4~~\\
\hline
\multirow{4}{*}{$[\bar{1}10]$}&$[001]_{+2\%}$&\multicolumn{2}{c}{0.306}&-0.02&-0.04&0.03&0.02&~~-0.1\\
&$[001]_{+5\%}$&0.307&0.306&-0.04&-0.07&0.07&0.04&-2\\ 
&[$\bar{1}10]_{+2\%}$&0.307&0.305&-0.04&-0.10&0.04&0.03&~-2~~\\
&[$\bar{1}10]_{+5\%}$&0.311&0.307&-0.05&-0.19&0.09&0.03&-16~~\\
\hline
\hline
\end{tabular}
\label{tab:umod}
\end{table}

In the remainder of this sub-section we now discuss the actual
displacement patterns of the atoms under different values of strain,
which will be used as reference for the atomic displacements within
the strained surfaces in Sec.~\ref{subsec:strained}. For this purpose,
we strain the bulk structure along the [001], respectively
$[\bar{1}10]$, direction and relax all atomic positions, whereas we do
not consider a further relaxation of the lattice vectors perpendicular
to the strain direction.\cite{hilfe} 
   If the atomic positions are
optimized within the metastable paraelectric state, the internal
parameter $u$ slightly increases with strain ($<$1~\% for imposed
strain of up to 5~\%), i.e., the ratio between the Ti-O$_{ap}$ and
Ti-O$_{eq}$ distances increases, as the Ti-O$_{eq}$ distances are less
sensitive towards strain or lattice expansion. This relaxation of the
atomic positions is in agreement with previous work, e.g.,
Ref.~\onlinecite{Montanari}.

If the paraelectric symmetry of the structure is lifted,
we obtain polar displacements of the atoms, in agreement to the
discussed softening of the polar modes under tensile strain, see
Table~\ref{tab:umod}. Polar displacements both along [001] and
$[\bar{1}10]$ directions lower the energy relative to the paraelectric
state under 2~\% and 5~\% tensile strain, and the cases with polar
displacements parallel to the strain direction are energetically most
favorable for both strain orientations, {\it{cf.}}
Ref~\onlinecite{Hong}.

We note that the amplitudes of the polar displacements are not the
same for all Ti/O atoms in case of [$\bar1$10] strain. Two
inequivalent Ti (Ti(1), Ti(2)) and O (O(1), O(2)) sub-lattices exist
under such strain, as the equatorial Ti(2)-O(2) bonds are not modified
while the imposed strain enlarges the apical Ti(2)-O bonds (the
opposite holds for the Ti(1)-O distances), see
Fig.~\ref{fig:rutil}~(b). As a result, the short-range repulsion and
thus the amplitudes of the atomic displacements differ for both atomic
subclasses. Specifically, the displacements along [001] are larger for
Ti(1)/O(1) than for Ti(2)/O(2) (and the other way round for the
[$\bar1$10] displacements), {\it{cf.}}
Refs.~\onlinecite{110,Bordeaux}.

\subsection{Born charges}
\label{subsec:born}
\begin{table}
\caption{ Contributions of individual MLWFs (see
  Fig.~\ref{fig:wannier}  and
  total $Z^*_{val}$. O$_{eq}$ (O$_{ap}$) correspond to MLWFs situated
  at equatorial (apical) oxygen atoms relative to the displaced
  Ti. Rows denoted ``Eq.''  are calculated for the equilibrium
  structure, rows denoted ``$u$+3\,\%'' are calculated for 3\,\%
  increased internal parameter $u$.  Upper part: Z$^*_{[001]}$; Lower
  part: Z$^*_{||}$. In each case the mean values for the corresponding
  orbitals are given.}
\begin{tabular}{lrrrrrrrr}
\hline
\hline
 & & \multicolumn{3}{c}{O$_{eq}$} & \multicolumn{3}{c}{O$_{ap}$} & $Z^*_{val}$ \\
 & & $p_\perp$ & $sp^2_{eq}$ & $sp^2_{ap}$ &  $p_\perp$ & $sp^2_{eq}$ & $sp^2_{ap}$ & \\
\hline
$Z^*_{[001]}$ & Eq. & 0.48 & 0.30 & 0.04 & $-$0.01 & $-$0.01 & $-$0.10 & 3.96 \\
      & $u$+3\,\% & 0.39 & 0.29 & 0.00 & $-$0.03 & $-$0.02 & $-$0.08 & 3.26 \\
\hline
$Z^*_{||}$ & Eq. & $-$0.36 & $-$0.04 & 0.06 & 0.48 & 0.23 & 0.29 & 3.38 \\
    & $u$+3\,\% & $-$0.23 & $-$0.08 & 0.05 & 0.53 & 0.27 & 0.31 & 4.20 \\
\hline
\hline
\end{tabular}
\label{tab:wannier}
\end{table}
\begin{figure*}
\fbox{
\includegraphics[width=0.23\textwidth]{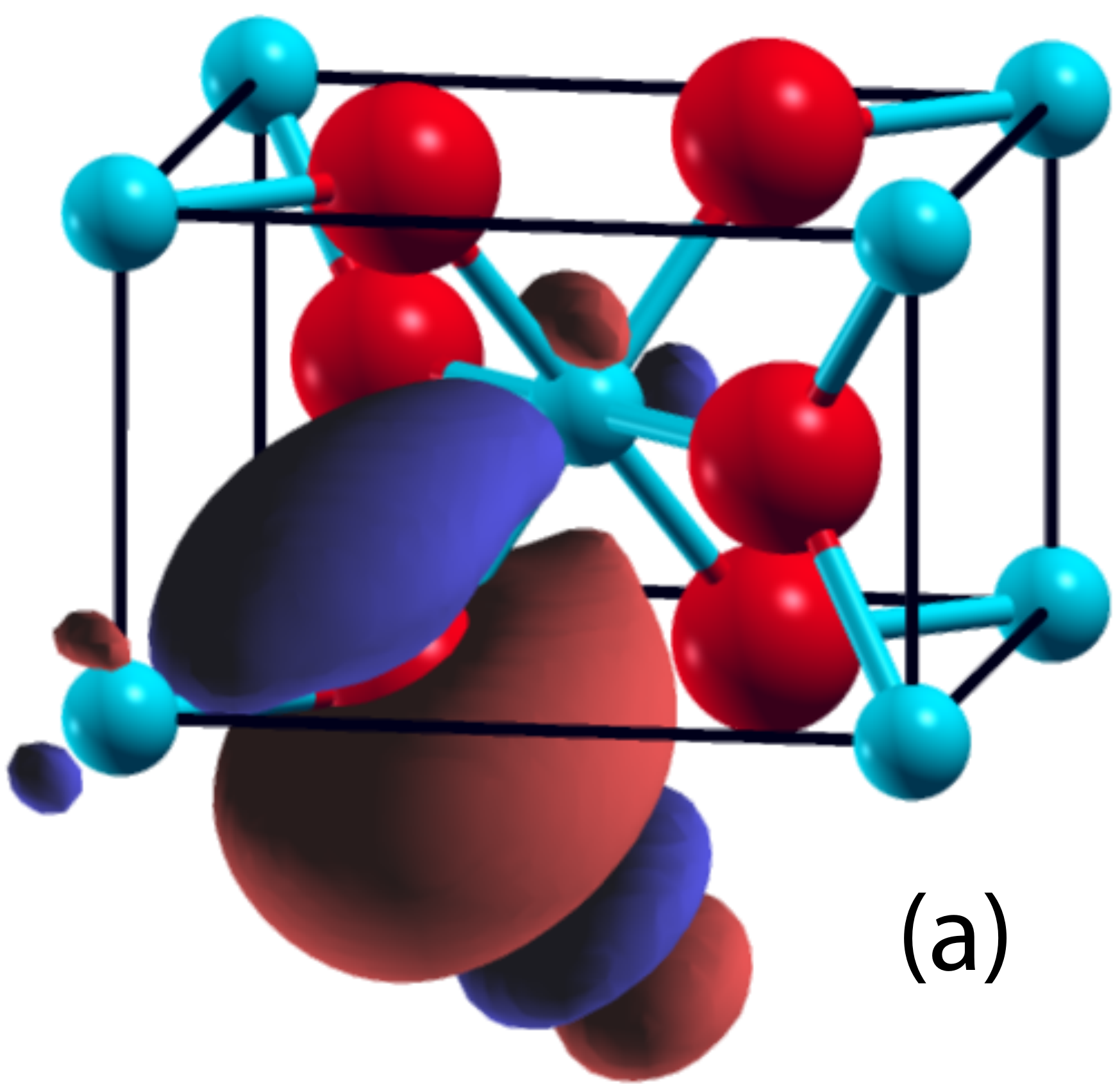}
\includegraphics[width=0.23\textwidth]{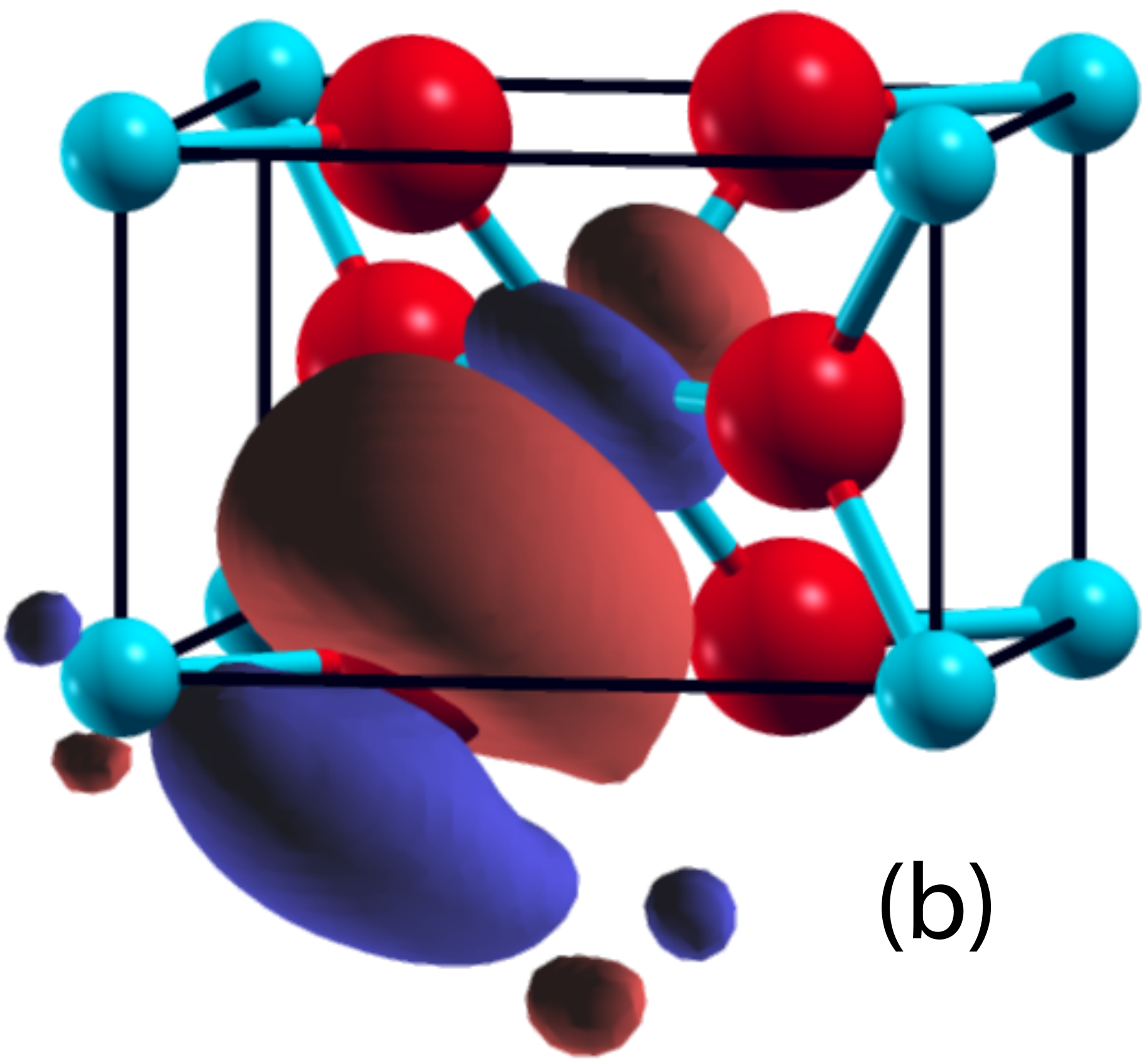}
\includegraphics[width=0.23\textwidth]{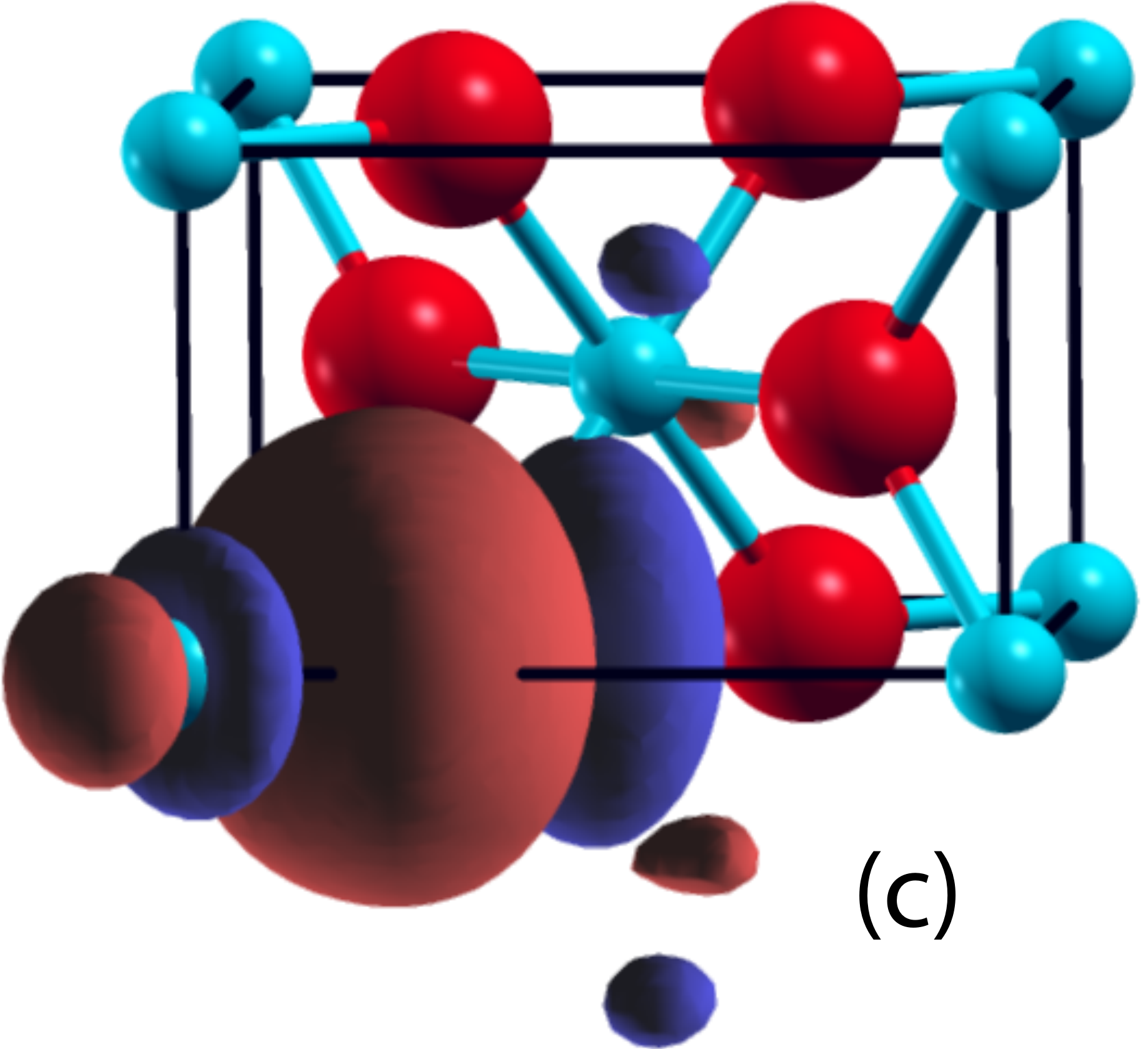}
\includegraphics[width=0.23\textwidth]{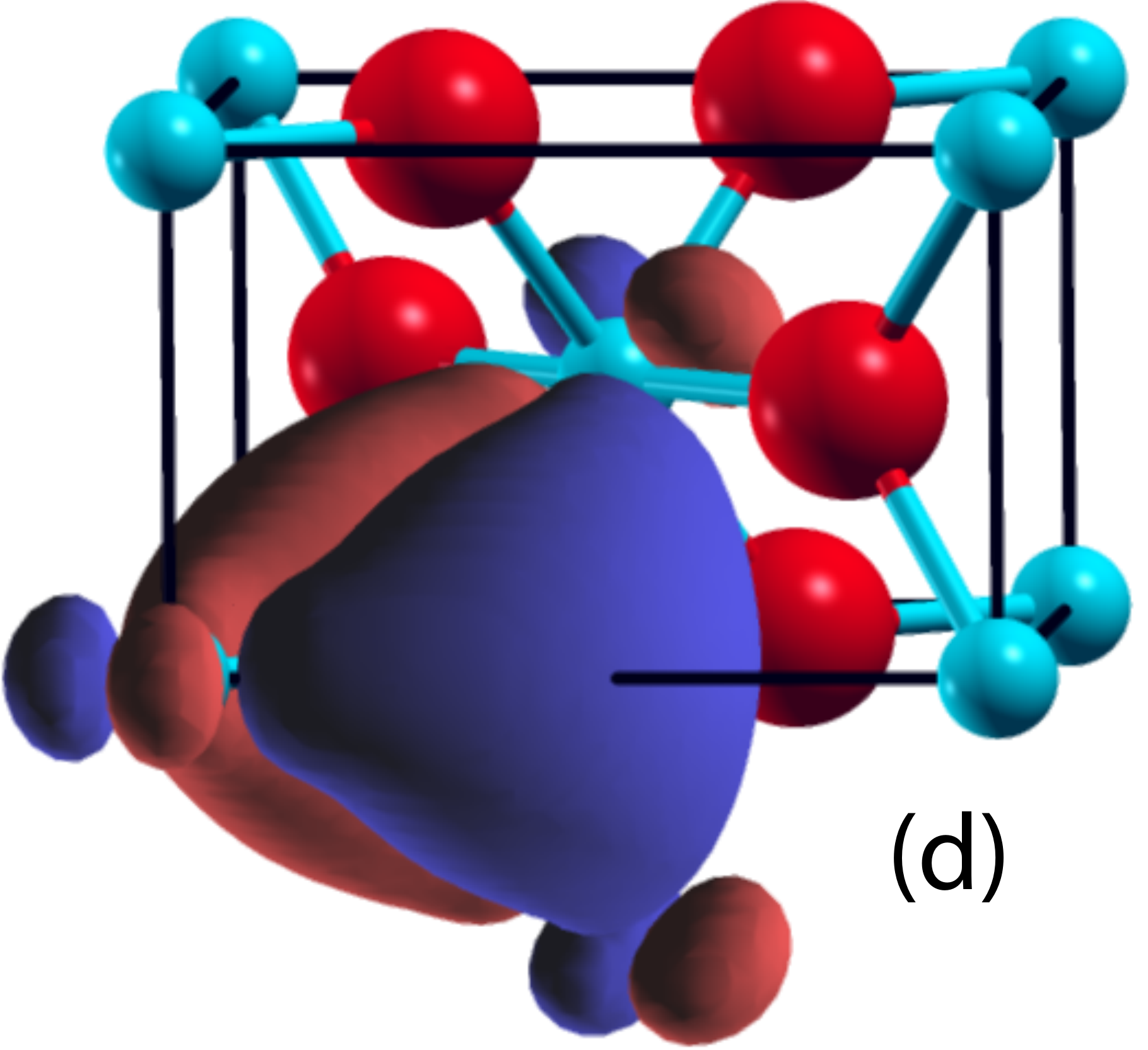}}
\caption{(Color online) Maximally localized Wannier functions corresponding to (a) $sp^2_{eq,dn}$,
 (b) $sp^2_{eq,up}$, (c) $sp^2_{ap}$ and (d) $p_{\perp}$-orbitals. }
\label{fig:wannier}
\end{figure*}

As already mentioned previously, the energy difference between the
paraelectric and the ferroelectric phase is often discussed in terms
of a delicate balance between short-range repulsive forces and the
long-range dipolar interaction which favor the ferroelectric
state.\cite{Cochran:1960} It has been further shown for perovskites
like BaTiO$_3$ that the dipolar interaction can counterbalance the
short-range repulsion if an anomalous hybridization between the
displaced ions lowers the energy during a ferroelectric
transition.\cite{Ghosez/Gonze/Michenaud:1996} The local chemical
environment in TiO$_2$ is quite similar to the perovskite structure,
with Ti$^{4+}$ ions that are octahedrally coordinated with O$^{2-}$
ions, and mostly ionic bonds with some covalent character. It is thus
very likely that similar dynamic hybridization effects as in
perovskites can also stabilize the ferroelectric state in strained
rutile. Indeed, a Mulliken population analysis showed an increased
Ti$_d$-O$_p$ hybridization within the ferroelectric
phase.\cite{Montanari}

To obtain a detailed understanding of this dynamic charge transfer, we
make use of the dynamic Born effective charge,
\begin{equation}
Z^*_{ij,\kappa}=\frac{\Omega}{|e|}\frac{\partial P_i}{\partial
  r_{\kappa,j}} \ ,
\end{equation}
which describes the change in polarization $P$ in direction $i$ for a
shift of atom $\kappa$ in direction $j$. Here, $\Omega$ is the unit
cell volume and $|e|$ the electronic charge. Due to the charge
neutrality condition, $2\cdot Z^*_{Ti}=-4\cdot Z^*_{O}$, we can
restrict the following discussion to $Z^*_{Ti}$. The effective charge
tensor of Ti in rutile is diagonal in a $[\bar{1}10]$, $[001]$,
$[110]$ reference system, see Fig.~\ref{fig:rutil}(b), with the
principal values $Z^*_{||}$, $Z^*_{[001]}$, and $Z^*_{\perp}$.  Here,
$Z^*_{||}$ is the dynamic charge along the apical bond, which is
pointing along $[110]$ respectively $[\bar{1}10]$ for every second
Ti-atom, and $Z^*_{\perp}$ is the charge along the third perpendicular
direction in each case.

In Tab.~\ref{tab:const} our values calculated with VASP and the values
obtained form the shifts of the center of gravities of the maximally
localized Wannier functions (MLWFs) based on PWscf calculations are
compared to literature.  Although the VASP results are slightly larger
than all other values, due to
the use of soft potentials, the qualitative trends are reproduced
within both approaches.

It can be seen that in particular the [001]-component of the Born
tensor, $Z^*_{[001]}$, is exceptionally large and of the same size as
for ferroelectric materials (BaTiO$_3$:
Z$^{Ti}_{[001]}$=7.3).\cite{Ghosez1} Furthermore, the Born charge along
the apical bond, $Z^*_{||}$, is also quite large and may stabilize a
ferroelectric transition in [110]/[$\bar{1}10]$ direction, whereas
Z$^*_{\perp}$ is only moderately enhanced compared to the nominal
ionic charge of $+4$.

As a first step towards the understanding of dipolar interactions at
the rutile surface it is essential to understand in more detail the
underlying mechanism for the exceptionally large Born effective charge
of Ti$^{4+}$ in bulk TiO$_2$, and how it is affected by different
lattice modifications. These aspects will therefore be discussed in the
remainder of this section.

\begin{figure}
\includegraphics[width=0.49\textwidth]{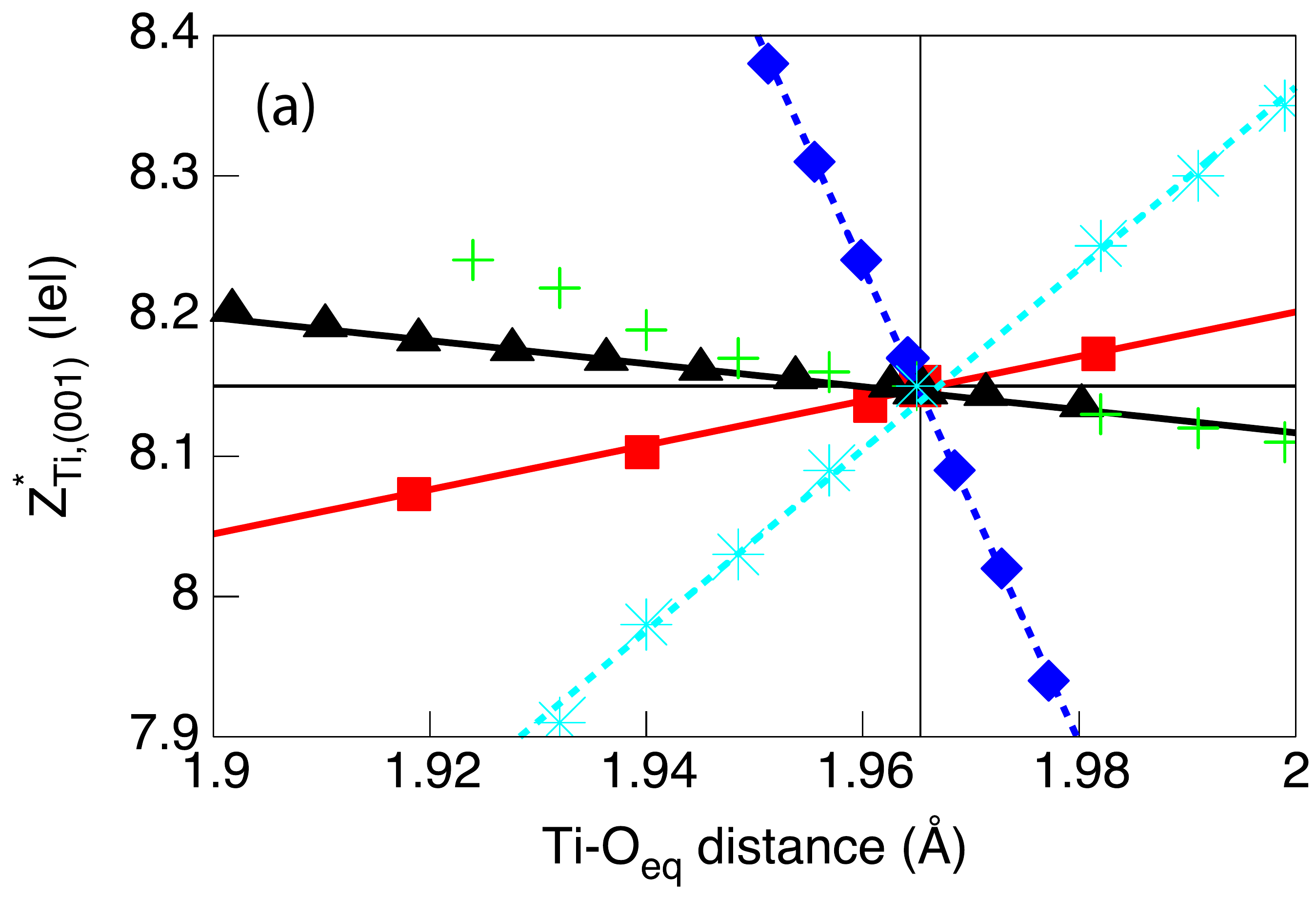}
\includegraphics[width=0.49\textwidth]{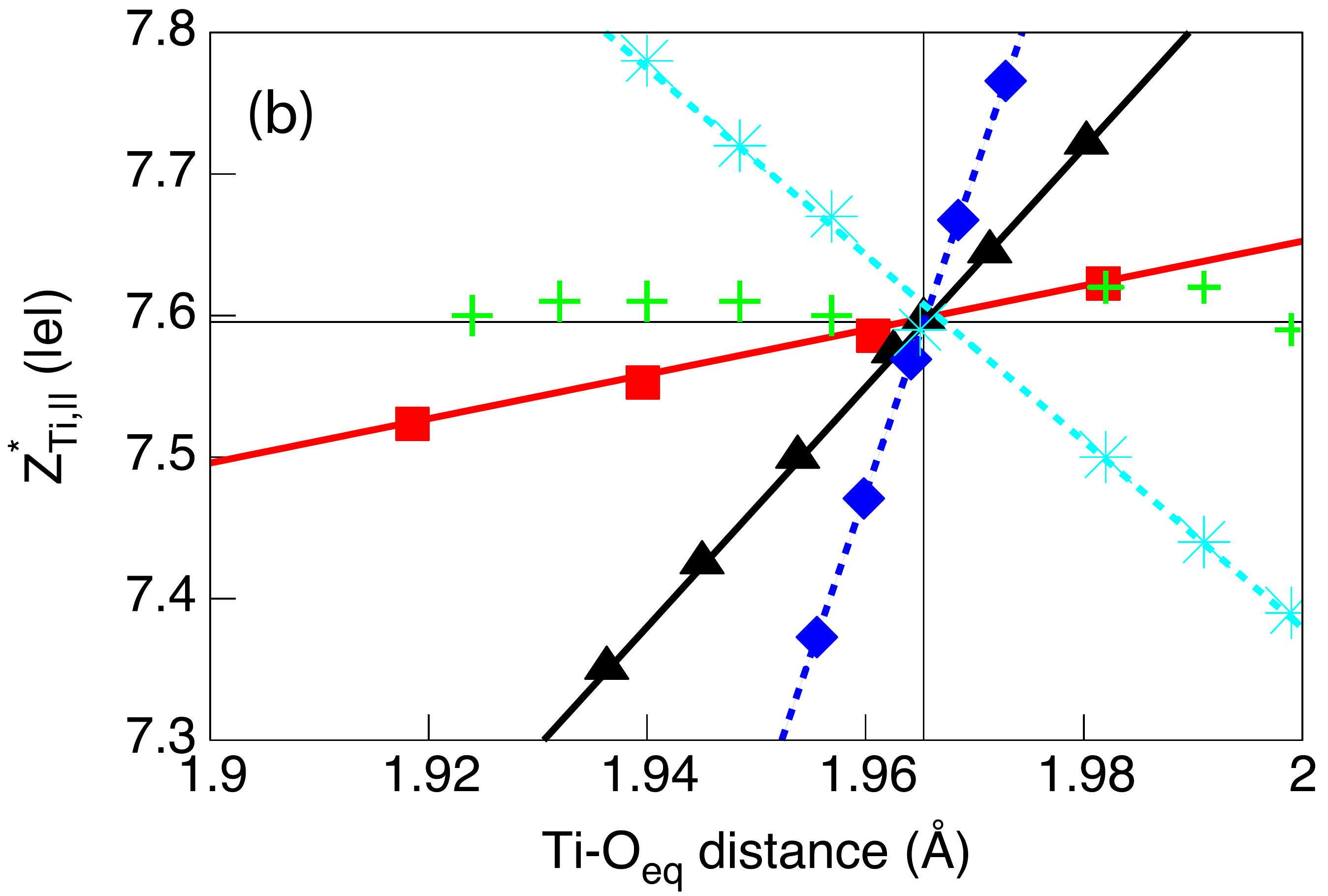}
\caption{(Color online) (a)-(b) Change in the Ti Born charge
  Z$^*_{[001]}$ (a) and Z$^*_{||}$ (b) due to a modification of the
  equatorial Ti-O bond length by uniform lattice expansion (red,
  squares), modification of $u$ (blue, diamonds), [001]-strain (black,
  triangles), and $[\bar{1}10]$ strain. For the case of $[\bar{1}10]$
  strain the given Ti-O$_{\text{eq}}$ distance corresponds to Ti(2)
  and the Born charges are depicted for both Ti(1) (cyan, stars) and
  Ti(2) (green, pluses).}
\label{fig:born}
\end{figure}

Based on the constructed MLWFs we decompose $Z^*_{Ti}$ according to
(see, e.g., Ref.~\onlinecite{Ederer}):
\begin{equation}
Z^*=Z^*_{core}+Z^*_{sc}+Z^*_{val} \ ,
\label{eq:ladung}
\end{equation}
where $Z^*_{core}=+12$ is the charge (in units of $|e|$) of the
nucleus screened by the inner electrons, $Z^*_{sc}$ the contribution
of the Ti $3s$ and $3p$ semi-core states, and Z$^*_{val}$ is the
contribution of the valence electrons. We find that the semi-core
contribution to $Z^*_{sc,[001])}$ ($Z^*_{sc,||}$) differs only by
$-0.23$ ($-0.12$) from the nominal value of $-8$. In contrast, large
anomalous values appear for Z$^*_{val}$, which can be further
decomposed into contributions of individual MLWFs, see
Tab.~\ref{tab:wannier}. Thereby, the 16 valence bands of TiO$_2$ with
dominant O$_{2s}$ and O$_{2p}$ atomic character are represented by 16
MLWFs per unit cell, 4 orbitals per O ion. Starting from projections
onto three atomic $sp^2$ hybrid orbitals, centered at each of the four
different oxygen atoms and oriented towards the three surrounding Ti
atoms, and one $p$ orbital oriented perpendicular to the corresponding
O-Ti$_3$ triangles (see Fig.~\ref{fig:rutil}(a)), we obtain three
$\sigma$ and one $\pi$-type orbital per oxygen as depicted in
Fig.~\ref{fig:wannier}. We note that depending on the initial
projection used for the spread minimization, different localized
Wannier orbitals can be obtained, corresponding to various local
minima of the total quadratic spread functional.
\begin{figure}
\includegraphics[width=0.5\textwidth]{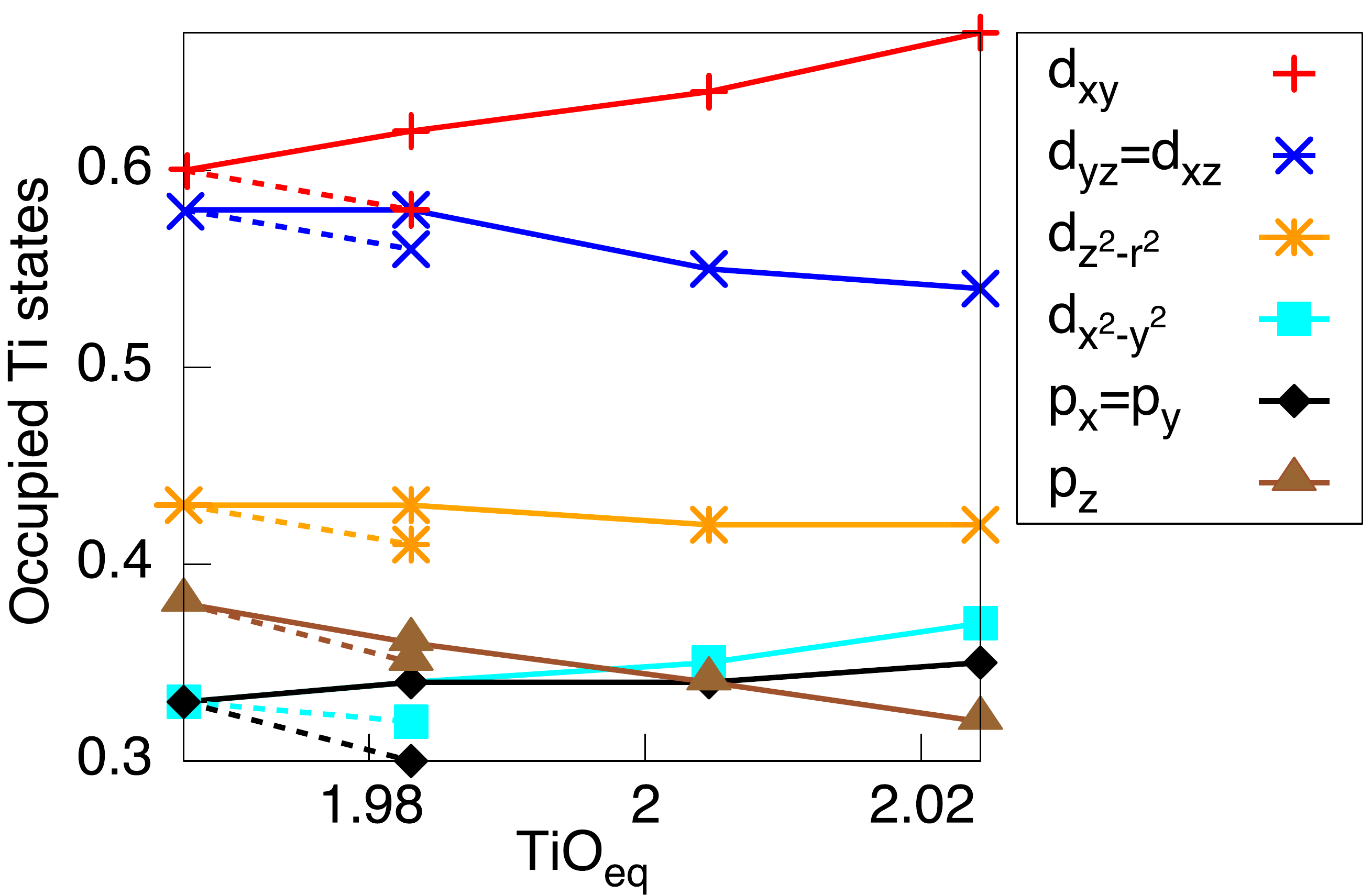}
\caption{(Color online) Orbitally-resolved changes in the integrated
  valence density of Ti states due to a modification of the Ti-O bond
  lengths. Solid lines: Modification of the internal parameter
  $u$. Dashed lines: Uniform lattice expansion. x: [001], y: [010], z:
  [001].}
\label{fig:hyb}
\end{figure}

In agreement with the formal ionic Ti$^{4+}$-O$^{2-}$ configuration,
the centers of gravity of each MLWF are close to the central
O-atoms. Nevertheless, the mixed covalent-ionic character of the Ti-O
bonds can clearly be seen from the $d$-character ``tails'' of the
MLWFs at the Ti positions (see Fig.~\ref{fig:wannier}). This is
consistent with the integrated projected density of states shown in
Fig.~\ref{fig:hyb}, which, for example, results in an occupation of
about 0.6 electrons for the Ti d$_{xy}$ atomic
orbital.\footnote{Projection of the total density of states onto
  atomic orbitals within spheres of 1.32~{\AA} for each Ti atom.}
Furthermore, the center of gravity of the $\sigma$-type MLWF along the
shorter equatorial bond divides the bond with a ratio of 22:78, while
the longer apical bond is divided with a ratio 20:80.  \C{Similarly,
  the tails of the $p$-type MLWFs are larger along the equatorial
  direction than along the apical.}

If the central Ti-atom is displaced along [001] ($[\bar{1}10]$), the
Ti-O$_{eq}$ (Ti-O$_{ap}$) distances are modified, and thus the
covalent character of the corresponding bonds change. This leads to a
shift of the center of gravity of the {MLWFs}\C{located on opposite to the
displacement of the Ti}, which causes large and positive anomalous
contributions to the Born effective charges, see
Tab.~\ref{tab:wannier} (we note that in the fully ionic case the
valence contribution to the Ti Born effective charge would be
identical to zero). Thus, charge is dynamically transferred in and out
of the Ti-$d$ orbitals. Similar to the case of the perovskite
ferroelectric BaTiO$_3$,\cite{Ederer} the $\pi$-type orbitals
($p_\perp$) show the largest anomalous contributions to $Z^*$,
underlining the large similarity between TiO$_2$ and the related
ferroelectric perovskites.

In addition to these positive contributions, the MLWFs located at
O$_{ap}$ (O$_{eq}$), exhibit small negative contributions to
$Z^*_{[001]}$ ($Z^*_{||}$). The corresponding Ti-O bond lengths do not
change to first order in the displacements and thus no strong change
in hybridization occurs. Instead, the charge contained in the
``tails'' of the corresponding MLWFs move with the Ti atom, leading to
small negative contributions to $Z^*$. This effect is especially
pronounced for the equatorial $\pi$-type orbital for a displacement of
the Ti atom along [$\bar{1}$10], thereby slightly reducing $Z^*_{||}$.


We note that a similar Wannier decomposition for the Born effective
charge of the oxygen anion in rutile TiO$_2$ has been presented in
Ref.~\onlinecite{Cangiani}. In agreement with our work, Cangiani
{\it{et al.}} found the largest anomalous contributions to $Z^*_O$ for
the $p_{\perp}$-type orbitals. However, even though the total charges
of Ti and O are related by the charge neutrality condition, the
relation between individual orbital contributions is less
straightforward, thus preventing a more detailed comparison.


One important factor for the modification of the dipolar interaction
at surfaces are atomic relaxation and structural distortions at the
surface. In the following we therefore consider different lattice
modifications within the bulk system and investigate the resulting
effect on the Born effective charges. Thereby we are considering
similar lattice modifications as in the previous section, i.e., a
uniform lattice expansion, modification of the ratio of the Ti-O
distances by $u$, and imposed strains in the [001] and [$\bar{1}$10]
directions, in each case without further relaxation of the other
parameters. The results are depicted in Fig.~\ref{fig:born}.

A uniform lattice expansion leads to a small increase of both
Z$^*_{[001]}$ and $Z^*_{||}$, while at the same time the static
hybridization of the atoms is reduced, see Fig.~\ref{fig:hyb}.
For all other distortions the ratio of the Ti-O bond lengths is
modified in addition to eventual changes in the volume. A variation of
the internal parameter $u$ in turn conserves the volume, while
increasing the equatorial bond distance and decreasing the apical bond
distance. As seen in Fig.~\ref{fig:hyb} this also changes the static
hybridization. On one hand, the occupation of the Ti-$d$ and $p$
states within the $x$-$y$ plane increases with the largest gain for
the $d_{xy}$ orbital, which is oriented along the apical bond. On the
other hand, the occupation of orbitals with lobes in [001] direction
is reduced. \C{, and especially the p$_z$-orbital is systematically
depopulated.} Thus, the Ti-O$_{eq}$ [Ti-O$_{ap}$] hybridization
decreases [increases], as do the weights of the corresponding tails in
the MLWFs. 

This change in the static hybridization seems to reduce the dynamic
change in hybridization between the Ti and O$_{eq}$ orbitals
especially for the $p_{\perp}$ orbital.  As a consequence, both the
positive and negative contributions of this orbital to Z$^*_{[001]}$
and Z$^*_{||}$, respectively, are reduced.  Similarly, the reduced
coupling of the $sp^2_{ap}$ orbital with the equatorial Ti neighbor
may explain the slight reduction of its contribution to Z$^*_{[001]}$.
In contrast, the hybridization between the Ti atom and its apical
neighbor increases for increased $u$, as do most of the negative and
positive contributions of these orbitals to Z$^*_{[001]}$ and
Z$^*_{||}$. In summary, a systematic reduction [increase] of the
effective charge along [001] ([110]) occurs, see
Fig.~\ref{fig:born}. For example, $Z^*_{val,[001]}$ [$Z^*_{val,||}$]
decreases [increases] by 8~\% [24~\%] within the PWscf calculation for
a $u$ modification of +3~\%. Thereby, the main reduction in
$Z^*_{val,[001]}$ is due to a reduced anomalous contribution of the
$p_{\perp}$ orbital at O$_{eq}$, whereas the increase of
Z$^*_{val,||}$ is due to slightly enlarged contributions of
$p_{\perp}$ and $sp^2_{eq}$ at O$_{ap}$ and a strong reduction of the
negative contribution of $p_{\perp}$ at O$_{eq}$.

One may wonder, whether the absolute Ti-O$_{ap}$ distance instead of
the Ti-O bond ratio plays the major role for this modification of the
static and dynamic hybridization. However, if the Ti-O$_{eq}$
distances are modified for a fixed Ti-O$_{ap}$ distance by a
combination of isotropic expansion and a variation of $u$, the
modification of $Z^*_{[001]}$ with the ratio of the Ti-O distances is
equal to a variation of $u$.

As the change of the Ti-O bond length ratio is smaller in the case of
[001] strain, the corresponding reduction [increase] of $Z^*_{[001]}$
[$Z^*_{val,||}$] is less pronounced. Similar to the modification of
Z$^*_{||}$, the third principle value of the Born tensor,
Z$^*_{\perp}$ slightly increases with a reduced ratio of the Ti-O
distances.  However, for small variations of the Ti-O distances, as
they may appear at the surface, the absolute value of Z$^*_{\perp}$ is
not significantly enhanced compared to the formal ionic charge, and
thus no considerable influence on the ferroelectric trends has to be
expected from this modification.

For $[\bar{1}10]$ strain, the ratio of Ti-O bonds corresponding to
Ti(2) is modified analogously to the case of [001] strain, and indeed
the corresponding decrease of $Z^*_{[001]}$ is identical to [001]
strain for small strain values, see Fig.~\ref{fig:born}(a). In
contrast, the apical bond length increases for Ti(1) while the
equatorial distance stays constant, and thus $Z^*_{[001]}$ increases
for this Ti atom. Furthermore, while $Z^*_{||}$ is nearly independent
of the equatorial bond distance for Ti(2) under $[\bar{1}10]$ strain,
$Z^*_{||}$ decreases drastically for Ti(1) with the corresponding
expansion of the apical bond.

In summary, the largest individual contributions to the anomalous Born
effective charges in rutile stem from the large dynamic charge
transfer along Ti-O $\pi$ bonds.  Additionally, the static Ti-O
hybridization, which depends strongly on the ratio of both Ti-O
distances, leads to considerable modifications of the effective
charges. The resulting modification of the dipolar interaction has a
considerable influence on the ferroelectric properties of the rutile
surface, as we will see in Sec.~\ref{subsec:unstrained}.

Furthermore, we find large effective charges which are stable against
small lattice distortions, as they may appear at the rutile surface.


\section{(110)-surface}
\label{sec:surface}

\begin{figure}
\includegraphics[width=0.5\textwidth]{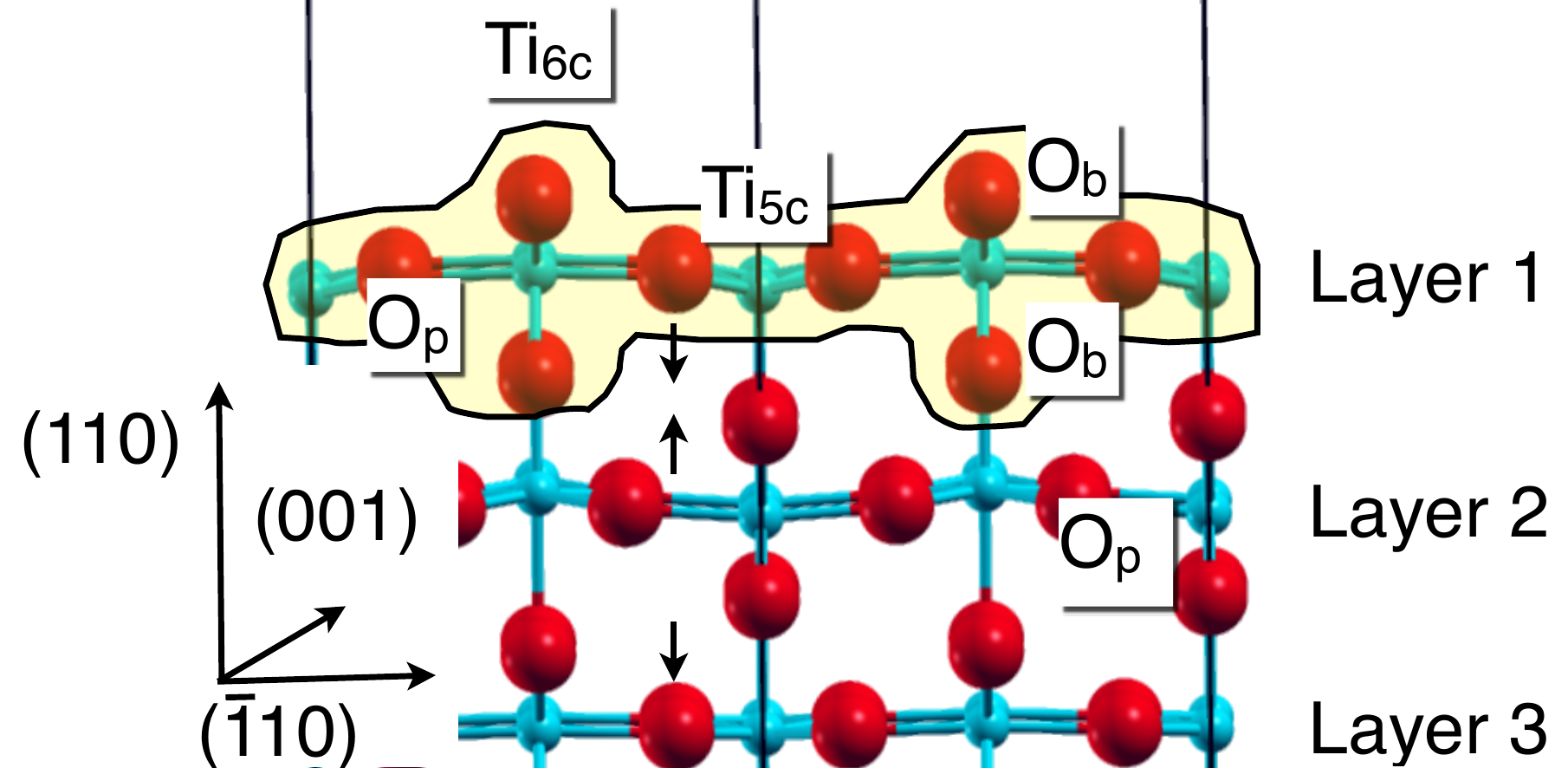}
\caption{(Color online) Atomic structure of the topmost layers of the
  rutile (110) surface. Light cyan: Ti, red: O. Black arrows:
  Interlayer distance oscillation.}
\label{fig:obstruk}
\end{figure}

\begin{table*}
 \setlength{\tabcolsep}{2pt}
 \caption{Ti-O bond lengths ({\AA}) within the three outermost layers
   of the rutile (110) surface within the paraelectric state (see
   Fig.~\ref{fig:obstruk}) with and
   without strain. The bulk distances refer to the optimized
   paraelectric structure with corresponding strain. Left columns:
   Ti$_{6c}$: sixfold coordinated Ti-atoms in the surface and Ti-rows
   below. Right columns: Ti$_{5c}$: fivefold coordinated Ti-atoms in
   the surface and Ti-rows below.  A single entry corresponds to Ti-O
   neighbors in plane while each Ti atom possesses different distances
   towards its inner/outer O$_b$ neighbors.}
\begin{tabular}{ccccccccccc}
\hline
\hline
&&bulk&&3&2&1&&3&2&1\\
\hline
\multirow{2}{*}{Eq.}&Ti-O$_{\text{eq}}$&1.97&\multirow{2}{*}{Ti$_{\text{6c}}$}&2.02/1.92&1.98&2.11/1.85&\multirow{2}{*}{Ti$_{\text{5c}}$}&1.96&1.90/2.04&1.96\\
&Ti-O$_{\text{ap}}$&2.01&&2.02&2.21/1.87&2.06&&1.92/2.13&2.00&1.83/-\\
\hline
\multirow{2}{*}{[001]$_{+2\%}$}&TiO$_{\text{eq}}$&1.96&\multirow{2}{*}{Ti$_{\text{6c}}$}&2.04/1.93&2.00&2.13/1.85&\multirow{2}{*}{T$i_{\text{5c}}$}&1.98&1.91/2.07&1.98\\
&TiO$_{\text{ap}}$&2.06&&2.03&2.20/1.86&2.06&&1.92/2.13&2.00&1.83/-\\
\hline
\multirow{2}{*}{[001]$_{+5\%}$}&TiO$_{\text{eq}}$&1.97&\multirow{2}{*}{Ti$_{\text{6c}}$}&2.07/1.94&2.04&2.16/1.87&\multirow{2}{*}{T$i_{\text{5c}}$}&2.01&1.92/2.10&2.01\\
&TiO$_{\text{eq}}$&2.01&& 2.03&2.18/1.84&2.07&&1.90/2.11&2.00&1.81/-\\
\hline
\multirow{2}{*}{[$\bar{1}10]_{+2\%}$}&TiO$_{\text{eq}}$&1.96&\multirow{2}{*}{Ti$_{\text{6c}}$}&2.00/1.92&2.00&2.09/1.85&\multirow{2}{*}{T$i_{\text{5c}}$}&1.97&1.90/2.03&1.97\\
&TiO$_{\text{ap}}$&2.06&&2.07&2.19/1.88&2.11&&1.94/2.12&2.04&1.85/-\\
\hline
\multirow{2}{*}{[$\bar{1}10]_{+5\%}$}&TiO$_{\text{eq}}$&1.97&\multirow{2}{*}{Ti$_{\text{6c}}$}&1.99/1.92&2.01&2.08/1.85&\multirow{2}{*}{Ti$_{\text{5c}}$}&1.98&1.90/2.01&1.98\\
&TiO$_{\text{ap}}$&2.01&&2.14&2.18/1.89&2.19&&1.96/2.12&2.11&1.86/-\\
\hline
\hline
\end{tabular}
\label{tab:strain}
\end{table*}

For the investigation of the ferroelectric trends of the rutile
surface, we use two different approaches.  After summarizing the
structural relaxations at the surface we first briefly discuss the
ferroelectric trends for a free film of 7 monolayers (ML), whereas in
Sec.~\ref{subsec:unstrained} and \ref{subsec:strained} we model the
clamping of the film due to an idealized substrate by fixing two
bottom layers to paraelectric, respectively ferroelectric positions of
the bulk material under different strain conditions (see
Tab.~\ref{tab:umod}).

As can be seen in Fig.\,\ref{fig:obstruk}, the (110) surface consists
of Ti-O$_p$ layers with bridging oxygen (O$_b$) atoms between these
layers. The surface Ti-atoms are alternately five-fold coordinated
(Ti$_{5c}$) and six-fold coordinated (Ti$_{6c}$).  We note that we use
the nomenclature Ti$_{6c}$ (Ti$_{5c}$) not only for the atoms in the
first surface layer, but also for the corresponding Ti atoms below.

\subsection{Surface relaxation}
\label{sec:surface-relaxation}

The calculated atomic structure of the paraelectric films is very
similar for free and clamped films and thus we restrict the discussion
to the structural relaxation of the clamped film. The corresponding
results are summarized in Table~\ref{tab:strain}. Even though the
rutile (110) surface is non-polar and stable without reconstruction,
the atomic structure is significantly modified at the surface.  It has
been discussed in literature that the atomic structure converges only
slowly with film thickness.\cite{Murugan,Bredow,Bates} However, we yield modifications of the obtained interatomic distances smaller than 0.01~{\AA} if the increase the film thickness to 9, only.
 Additionally, our obtained structural relaxation
at the surface agrees well with previous results in
literature.\cite{Murugan,Bredow,Bates}

Most notably, the outermost Ti$_{5c}$ and O$_b$ atoms in the first
surface layer relax inwards and reduce their distance towards the
underlying apical and equatorial neighbors by 9~\%, respectively
6.1~\% (6.6~\%), in our calculations (in Ref.~\onlinecite{Bates}). In
contrast, the next deeper Ti-O bonds along the surface normal are
enlarged by 3.6~\%, respectively 7.1~\%, for the equatorial
Ti$_{5c}$-O$_b$ and Ti$_{6c}$-O$_b$ pairs. This alternating
modification of the Ti-O bonds reaches several {\AA} deep into the
surface, {\it{cf.}}  Refs.~\onlinecite{Murugan, Bredow}.

Apart from this oscillation of the Ti-O distances, a small surface
buckling appears, as Ti$_{5c}$-atoms and bridging O-atoms (O$_b$)
relax inwards whereas Ti$_{6c}$ and O atoms within Ti-O planes (O$_p$)
relax outwards, see Fig.~\ref{fig:obstruk}. It has been debated in
literature\cite{Bickel} whether an alike polar surface relaxation of
SrTiO$_3$ can be interpreted as ferroelectric relaxation along the
surface normal. However, this relaxation does not break the symmetry of
the ideal surface and thus cannot be interpreted as a ferroelectric
transition, nor exist two switchable states.\cite{Padilla}

In summary, the Ti-O bonds along the surface normal are considerably
modified, and show a large anisotropy for each Ti-atom, whereas the
bond lengths perpendicular to the surface normal are very close to
their bulk values, see Tab.~\ref{tab:strain}.

\subsection{Free surface}
\label{subsec:free}
Due to the strong symmetry lowering at the surface, the ferroelectric
phonon modes of the bulk are no longer eigenmodes of the system.  Most
notably, the Ti-O$_{eq}$ distances vary between 1.83~{\AA} and
2.11~{\AA}, see Tab.~\ref{tab:strain}. This means that the short-range
repulsion during a ferroelectric shift is locally very different, and
one may expect different amplitudes of polar displacements in each
layer. As a result of this, a polar displacement of the whole slab
along the bulk A$_{2u}$ mode (i.e., with the same relative
displacement between Ti and O in all layers) is energetically very
unfavorable. The stiffness of such a displacement increases by a
factor of ten in comparison to bulk.  

However, if all atomic positions are optimized after such a static
displacement along the bulk A$_{2u}$ mode, small polar displacements
persist within the film. For instance, the total energy of the
paraelectric system is lowered by 3 meV/cell if mainly the Ti$_{6c}$
atom in layer 3 is displaced by 0.07~{\AA} along [001] against its
surrounding O octahedra, resulting in a dipole moment of
0.8~$|e|\cdot${\AA} (determined with Eq.~\ref{eq:dipol}). For this Ti
atom, the enlarged short-range repulsion due to the shrink of two
equatorial bonds of 2~\% can be compensated, as the apical bond is
nearly conserved and the two other equatorial bonds increase by about
3~\%.  A ferroelectric configuration becomes even more favorable if
small polar displacements in $[\bar{1}10]$ direction are superimposed
to the displacement along [001], and we obtain an energy gain of 5
meV/atom relative to the paraelectric surface. 

Although these small energy well depths will most likely be overcome
by thermal or even quantum fluctuations, the presence of such polar
surface states confirms the strong ferroelectric trend at the rutile
surface, which is comparable to the SrTiO$_3$ surface.\cite{Padilla}
In addition, these polar displacement patterns demonstrate the large
modification of the polar phonon mode at the surface. Thus, in order
to investigate the ferroelectric trends of the free surface using the
frozen phonon approach (similar to what has been done for the bulk
material, see Sec.~\ref{subsec:modes}), it would be necessary to
calculate mode eigenvectors for the strained films. However, since the
determination of surface phonon modes is computational very demanding,
we will follow a slightly different approach in order to investigate
the ferroelectric trends of the surface in the remainder of this
paper.

\subsection{Clamped surfaces}

\label{subsec:unstrained}
\begin{figure*}
\includegraphics[width=0.4\textwidth]{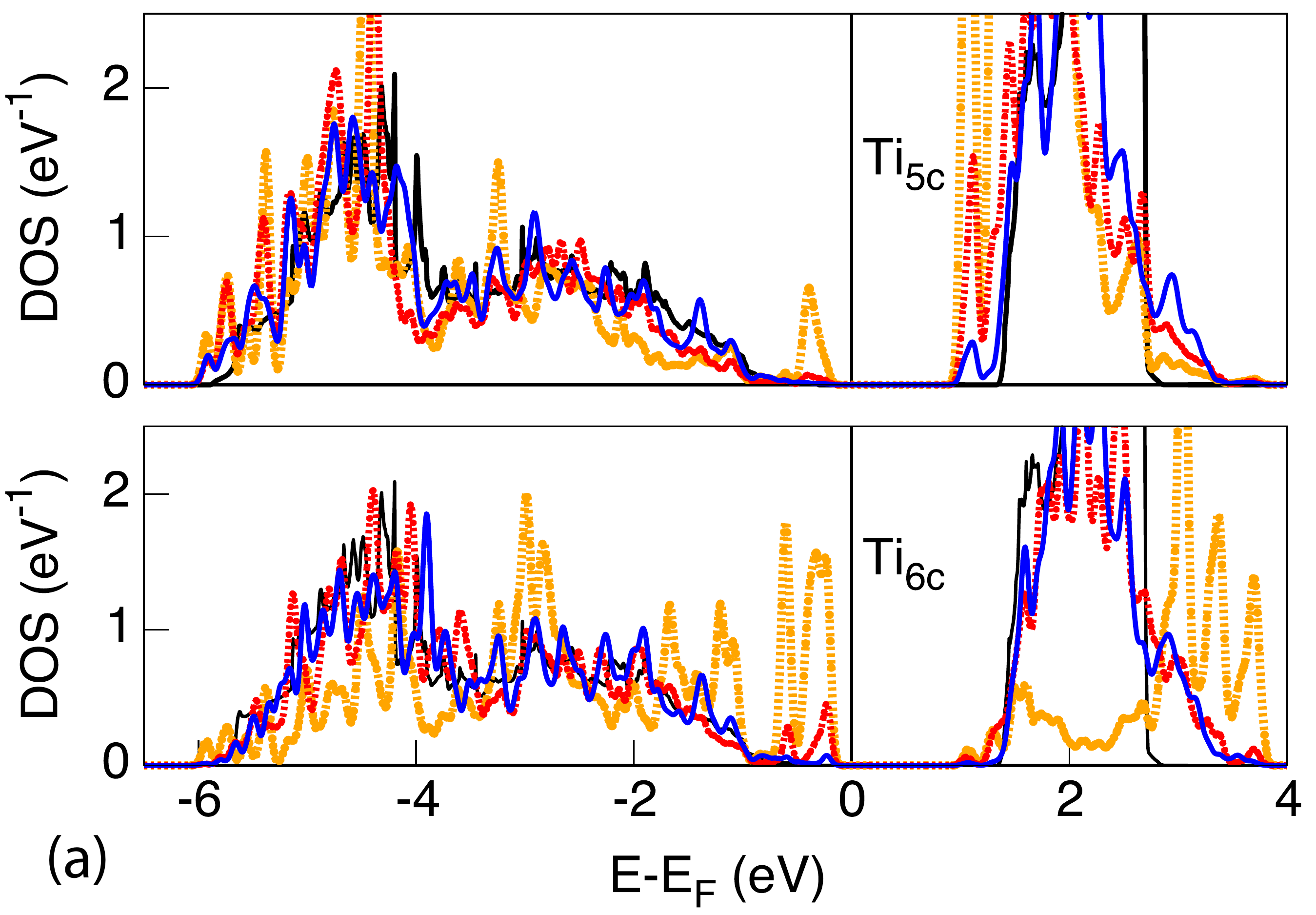}
\includegraphics[width=0.4\textwidth]{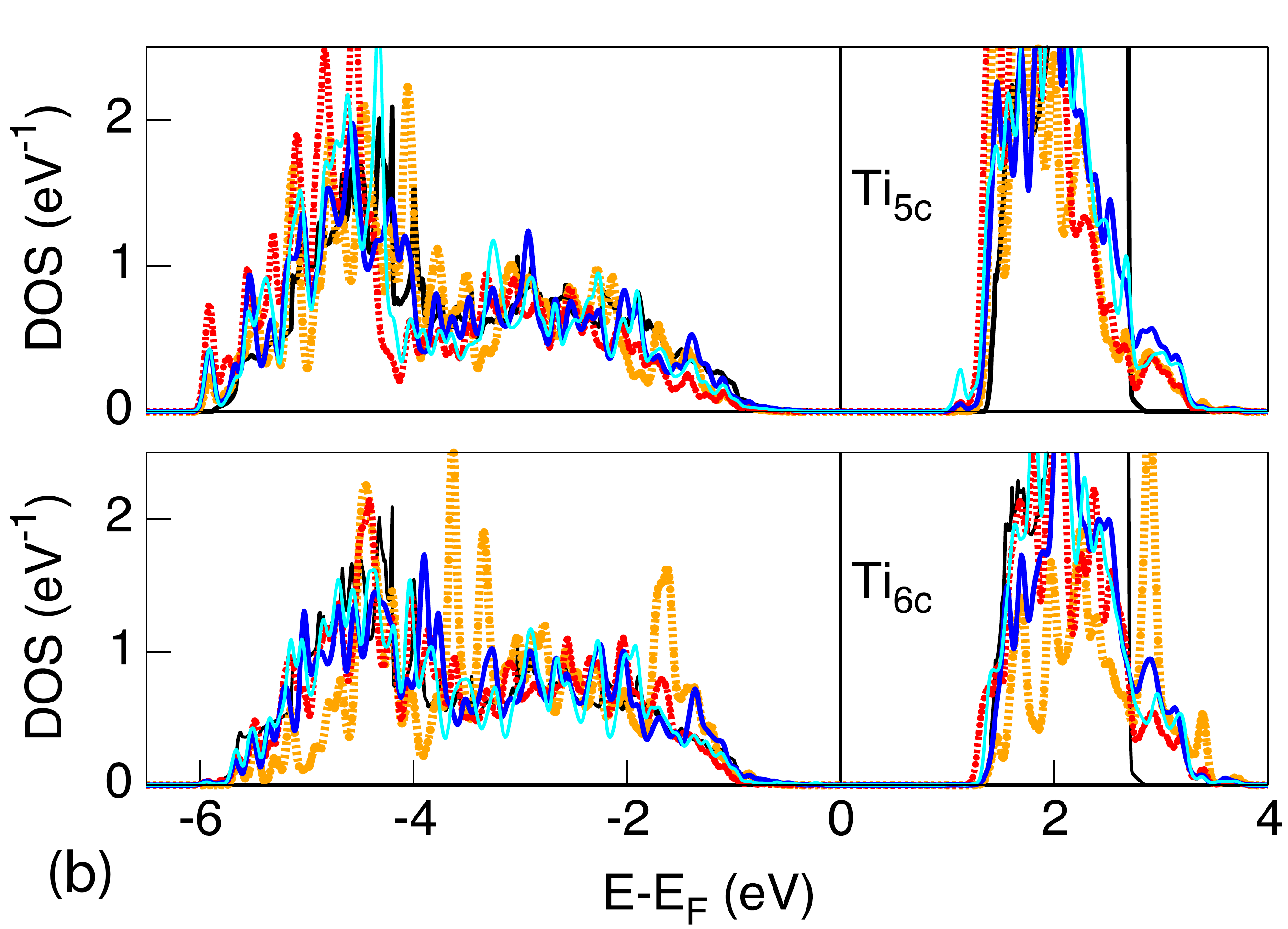}
\caption{(Color online) Layer resolved Ti density of states: 1st layer
  (yellow, light, dotted); 2nd layer (red, darker, dotted); 3rd layer
  (blue, dark, solid line); 4th layer (cyan, light, solid line) and
  bulk value as thin black lines. (a) The numbering starts at the
  fixed bottom of the film (b) The numbering of layers starts at the
  relaxed surface. \C{(c) Low lying Ti-O hybrid states for the topmost
  atoms within the relaxed surface. $d$-states along the surface
  normal (green, thick solid line) and $p$-states (magenta, thin solid
  line).}}
\label{fig:doslayer}
\end{figure*}

We use the atomic structure of the bulk material within the
paraelectric or ferroelectric phases, see Tab.~\ref{tab:p}, as
starting point and fix two bottom layers to these bulk positions. This
setup corresponds to a clamping of the bottom layers to an underlying
substrate. By using this procedure, the amplitude of the displacements
within the upper layers can account for the local atomic arrangement
of the surface, and thus will approach the real eigenstate of the
system. 

To quantify the local polar displacements, we use the total dipole
moment $p_i$ per Ti$_2$O$_4$ layer $i$:
\begin{equation}
\vec{p_{i}}=\sum_j Z^{*j}_{\vec{r}} \Delta \vec{r}_{j} \ ,
\label{eq:dipol}
\end{equation}
where $Z^{*j}$ is the Born charge of atom $j$ along $\vec{r}$, and
$\Delta \vec{r}_{j}$ is the shift with respect to the paraelectric
reference configuration of each atom $j$ within layer $i$.

As mentioned previously, the surface properties of TiO$_2$ converge
only slowly with film thickness. This is especially important for
electronic properties such as the band gap.\cite{Murugan, Bredow}
Furthermore, the clamping of one surface of the slab could induce
artificial electronic surface states which could then modify the
ferroelectric trends. To address this issue, Fig.~\ref{fig:doslayer}
shows the layer-resolved total density of states (DOS) corresponding to the
Ti atoms. Indeed,
surface states appear within the band gap at the fixed bottom surface,
see Fig.~\ref{fig:doslayer}(a). However, these states decay rapidly
with increasing distance from the bottom layer and do not
significantly influence the electronic structure of the free top
surface.
In addition, the DOS obtained here is in qualitative agreement with
the electronic structure obtained in previous
investigations.\cite{Bredow} The largest modification of the
electronic structure in comparison to the bulk material is a slight
narrowing of the band gap, due to a small shift of the unoccupied
$d$-states of the Ti$_{5c}$ atoms.  

\begin{figure*}
\includegraphics[width=\textwidth]{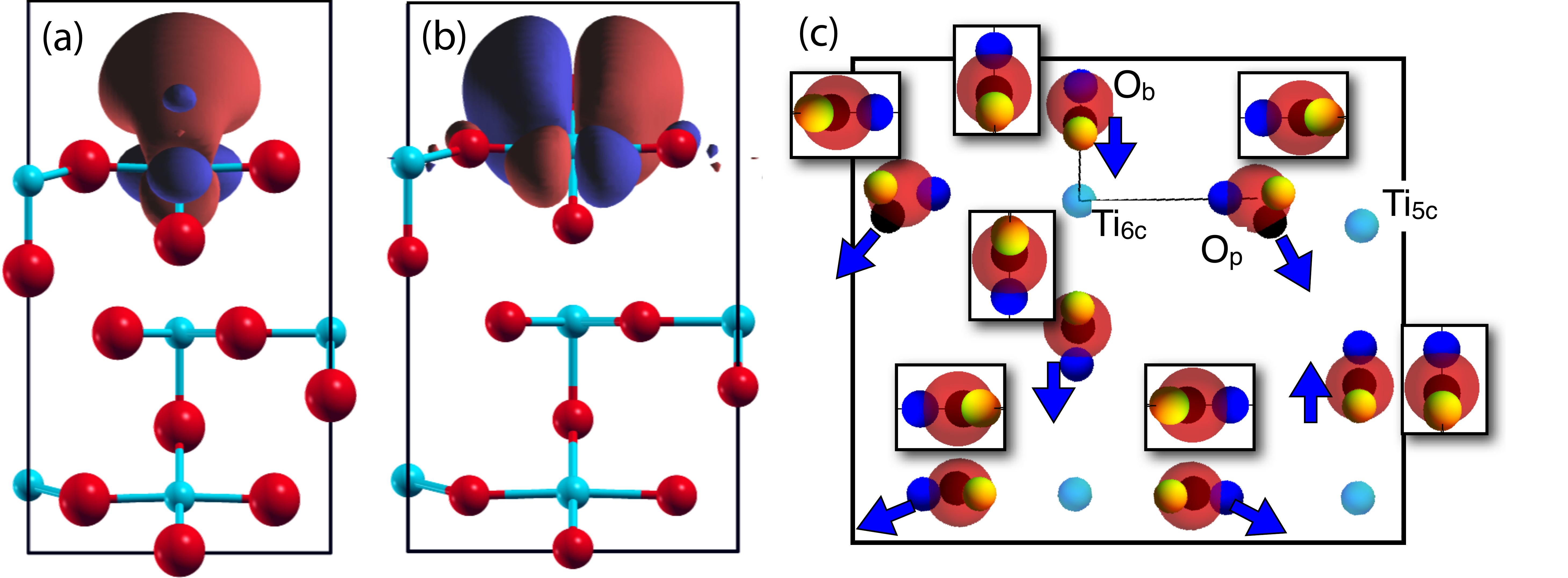}
\caption{(Color online) (a)-(b) MLWFs of the topmost O$_b$ atom. (a)
  sp$^2_{ap}$ orbital (b) p$\perp$ orbital. (c) Centers of gravity of
  the MLWFs close to the surface. Light, yellow spheres: sp$^2_{eq}$;
  dark, blue spheres: sp$^2_{ap}$; black spheres:
  $p_\perp$-orbital. The bulk distribution of the MLWF centers are
  given as reference within the small boxes. Arrows mark the overall
  shift of charge at the surface relative to the bulk structure.}
\label{fig:wannierob}
\end{figure*}

As discussed by Bredow {\it et al.},\cite{Bredow} the hybridization
between Ti and O states which are oriented along the surface normal,
oscillates with the surface distance. These surface-induced
modifications of the hybridization are correlated with the reduced
(enlarged) Ti-O distances at the surface, see
Sec.~\ref{sec:surface-relaxation}. As we showed in
Sec.~\ref{subsec:born}, small changes in the Ti-O arrangement and
hybridization can have considerable influence on the ferroelectric
characteristics. Because of this, it is essential to have a detailed
understanding of the character of the Ti-O bonding in the vicinity of
the surface in order to make predictions on the ferroelectric
trends. For this purpose, we decompose the valence states of the
surface slab into {MLWFs}, see Fig.~\ref{fig:wannierob}. We note that
in order to obtain qualitative trends at a reasonable computational
effort, this decomposition has been performed for a free film with a
thickness of only three monolayers.

It can be seen from Fig.~\ref{fig:wannierob}(a)-(b) that the MLWFs
with $sp^2_{ap}$ and $p_\perp$ character which belong to the topmost
O$_b$ atom are considerably modified in comparison to bulk. For both
orbitals, the Ti-atom in the apical position is missing at the
surface, and the centers of gravity shift towards the Ti$_{6c}$ atom
below, see subfigure (c). Thus, the covalent character of this bond
increases. Similarly, the hybridization of the undercoordinated
Ti$_{5c}$ surface atom and the O$_{b}$ atom below is
enlarged. Overall, an alternating increase and decrease of the Ti-O
hybridization appears along the surface normal, and the amplitude of
this oscillation decreases with the surface distance, fully consistent
with the changes in the corresponding bond lengths. Thereby, the
strongest shifts correspond to the centers of gravity of the
p$_\perp$-type {MLWFs}, which, as shown in Sec.~\ref{subsec:born},
also exhibit the largest individual contributions to the bulk Born
effective charges. Furthermore, we also found in Sec.~\ref{subsec:born}
that a modification of this p$_\perp$ orbital also leads to a
considerable modification of the dynamic charges, and thus an
oscillating modification of the dynamic charges can be expected at the
rutile surface.

\begin{figure}
\includegraphics[width=0.5\textwidth]{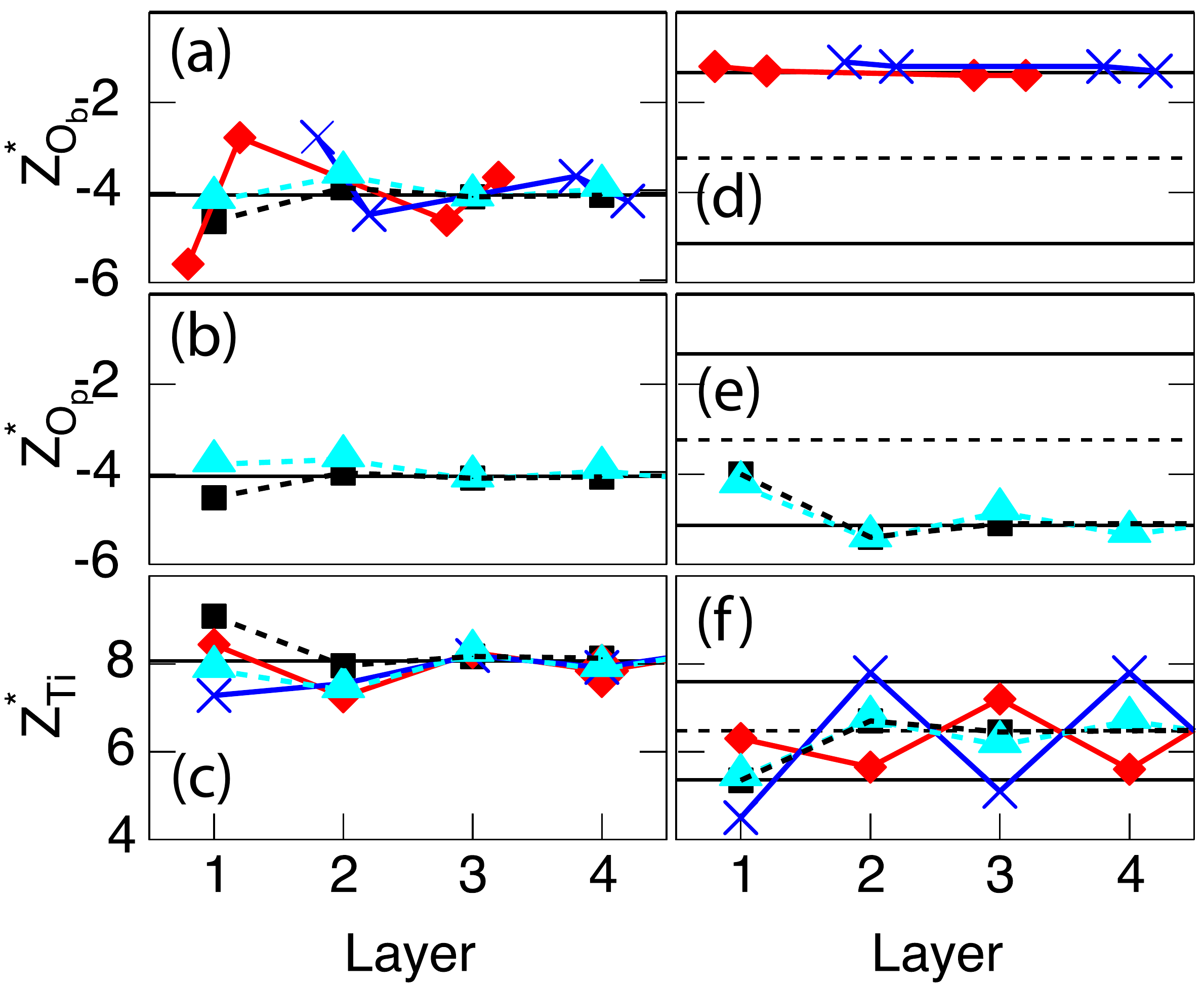}
\caption{(Color online) Layer-resolved Born charge at the rutile (110)
  surface. (a)-(c) Z$^*_{[001]}$ (d)-(e) Z$^*_{[\bar{1}10]}$.  Bulk
  values of Z$^*_{[001]}$, Z$^*_{||}$ and Z$^*_{\perp}$, (the mean
  value of Z$^*_{[\bar{1}10]}$) are marked by solid (dotted) lines.
  (a)/(d): O$_b$ (b)/(e): O$_p$, (c)/(f): Ti; \\Ti$_{6c}$ atoms and
  the corresponding O$_{eq}$ atoms (red, diamonds); Ti$_{5c}$ and the
  corresponding O$_{eq}$ atoms (blue, cross); mean value (cyan,
  triangles); mean value of the unrelaxed cell (black, squares). Lines
  are a guide to the eye only.}
\label{fig:bornlayer}
\end{figure}

In order to verify this conclusion, we now discuss the modifications
of the Born effective charges at the (110) surface in more detail.
Fig.~\ref{fig:bornlayer} shows the Born charges along [001] and
[$\bar{1}$10] for each class of atoms.\footnote{Due to the reduced
  symmetry at the surface, small off-diagonal elements of the Born
  tensor appear, which we neglect in our discussion.} We only consider
cases with polarization within the surface plane, in order to avoid
complications due to the strong depolarizing fields perpendicular to
the surface. The Born charges are calculated both for the structurally
relaxed as well as for the unrelaxed surface configuration.

For a detailed discussion of dipolar interactions along [$\bar{1}$10],
one has to keep in mind that Z$^*_{||}$ and Z$^*_{\perp}$ are
alternating in this direction. While all apical bonds are aligned
along [110] for O$_b$ (Z$^*_{[\bar{1}10]}$=Z$^*_{\perp}$), they are
aligned within the surface for the O$_p$ atoms
(Z$^*_{[\bar{1}10]}$=Z$^*_{||}$). For the Ti-atoms, Z$^*_{||}$ and
Z$^*_{\perp}$ are alternating in each layer, see
Fig.~\ref{fig:bornlayer}(d-f).

It is apparent from Fig.~\ref{fig:bornlayer} that even if atomic
relaxations are neglected, the dynamic charges are strongly modified
within the topmost layers. In particular,
Z$^{*}_{[001]}$ of the outermost Ti$_{6c}$ [(a)] and O$_b$ atoms [(c)]
is significantly enlarged in comparison to bulk, while the
corresponding charges within the topmost Ti$_{5c}$ [(a)] atom and its
equatorial O$_p$ [(b)] neighbors is slightly reduced.  Also, the
dynamic charges along [$\bar{1}$10] are reduced within the topmost
surface layer, as the mean value Z$^*_{[\bar{1}10]}$ of the Ti-atoms
[(f)] and Z$^*_{[\bar{1}10]}$ of O$_p$ [(e)] atoms are slightly
reduced.

If we account for atomic relaxations, the alternating strong and weak
Ti-O bonds appear along the surface normal, as discussed
above. Especially, the modification of the Ti-O$_{eq}$ bonds has a
considerable effect on the Born charges along [001].  Simultaneous
with the modification of the equatorial Ti-O$_{b}$ bonds,
Z$^*_{[001]}$ of these O$_b$ atoms oscillates, i.e. the dynamic charge
transfer increases (decreases), if the corresponding Ti-O$_{eq}$ bonds
are strengthened (weakened), while the amplitude of this oscillation
decreases with increasing distance from the surface (as does the
oscillation of the bond strength).  This is fully consistent with the
reduction (increase) of Z$^*_{[001]}$ for a reduced (enlarged)
Ti-O$_{eq}$ static hybridization of the bulk material, see
Sec.~\ref{subsec:born}

We note that the charge neutrality condition does not hold for each
layer under the reduced symmetry at the surface, and as a result about
0.2 electrons are transferred from the dynamic charges of layer $1$ to
the ones of layer $2$.  In particular, the Born charges of the
uppermost Ti$_{5c}$ atom and the Ti$_{6c}$ atom in layer 2 are not
balanced by their equatorial neighbors. For these Ti atoms, the
Ti-O$_{ap}$ bonds along the surface normal are reduced to 1.83~{\AA},
respectively 1.87~{\AA} and are thus shorter than the bulk Ti-O$_{eq}$
bonds.  Because of this, the Born charge in layer 1 is slightly
reduced.  Besides the direct surface, the mean Born charges per layer
of Ti, O$_p$ and O$_b$ atoms oscillate with a slight increase in odd
layers.

The mean values of Z$^*_{[\bar{1}10]}$ also oscillate with the surface
distance for Ti and O$_p$ atoms, see Fig.~\ref{fig:bornlayer}(e)-(f),
but, the Born charge is enlarged within even layers opposite to
Z$^{*}_{[001]}$.  This behavior is consistent with the opposite trends
of Z$^*_{[001]}$ and Z$^*_{||}$ for different modification of the Ti-O
bond ratio and hybridization in the bulk.  Z$^*_{\perp}$ of the O$_b$
atoms is nearly constant while Z$^*_{\perp}$ of Ti$_{6c}$ (Ti$_{5c}$)
increase (decrease) marginally.  As the Ti-O bonds within the surface
planes are modified to a smaller extent, the oscillations of
Z$^*_{||}$ of O$_p$ atoms are only very minor.  Besides this,
Z$^*_{||}$ of Ti$_{6c}$ (Ti$_{5c}$) is reduced (enlarged) in all
layers and a dynamic charge transfer of about 0.2 $|e|$ appears from
layers with odd numbers to layers with even number.\\ \C{Furthermore,
  the in-plane modification of the Ti-O$_{eq}$ bonds is much smaller,
  as is the change of the dynamic charge transfer within the planes.}
Overall, besides the discussed modifications of the Born effective
charges, the large anomalous dynamic charges along [001] and
[$\bar{1}10$] found in the bulk system are retained at the (110)
surface of rutile. Therefore, a similar tendency towards a
ferroelectric distortions as in the bulk material can also be
expected at the surface.

\subsection{Strained surface}
\label{subsec:strained}
\begin{figure}
\includegraphics[width=0.45\textwidth]{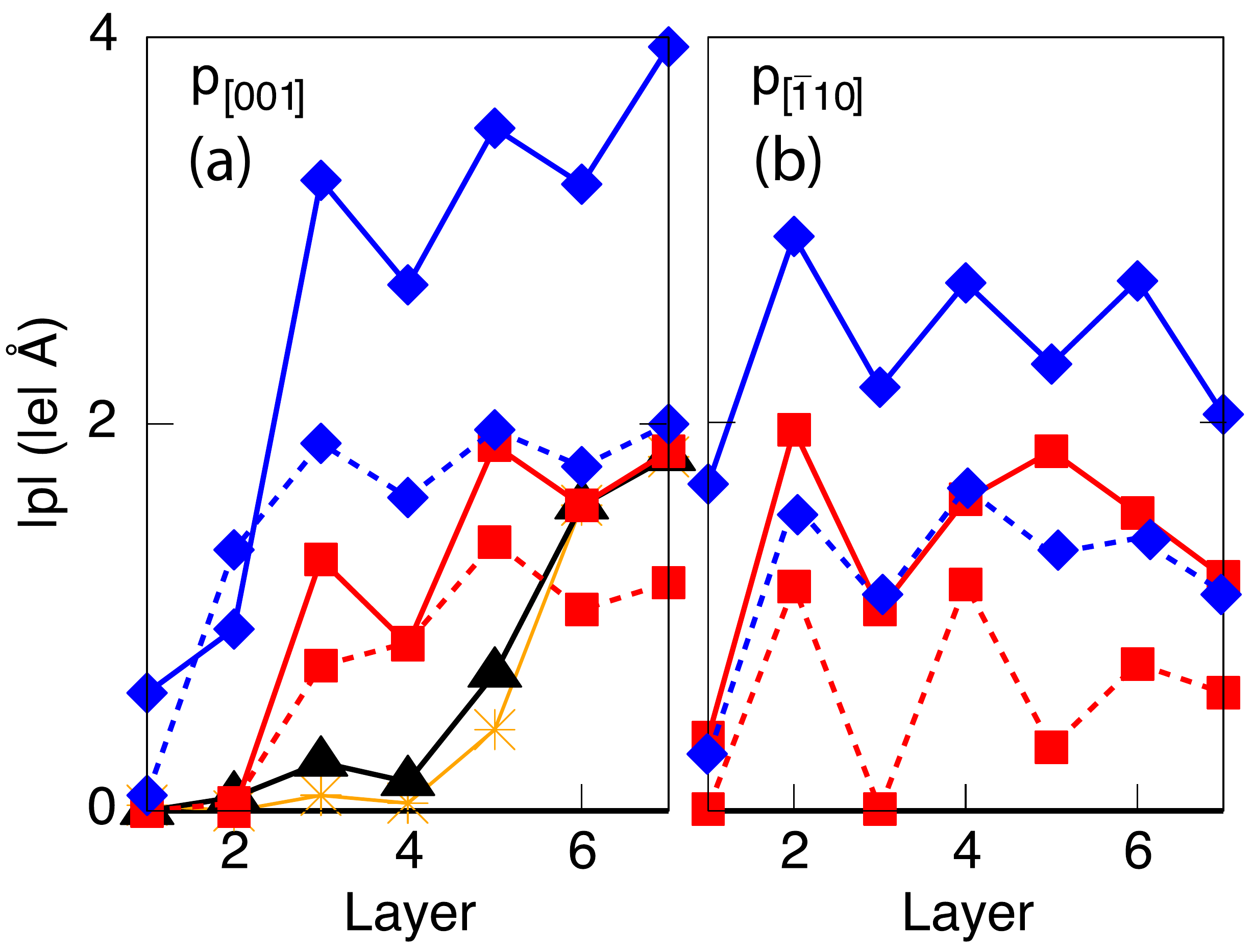}
\caption{(Color online) Layer-resolved dipole moments under tensile
  uniaxial strain. (a) $|\vec{p}_{[001]}|$ (b)
  $|\vec{p}_{[\bar{1}10]}|$.  Solid (dotted) lines correspond to
  strain parallel (perpendicular) to the polarization. 2~\%
  compressive strain (yellow, stars); no strain (black, crosses); 2~\%
  tensile strain (red, squares); 5~\% tensile strain (blue,
  diamonds). In all cases polar distortions perpendicular to the
  polarization direction of the substrate are prevented by the imposed
  symmetry.}
\label{fig:dipole}
\end{figure}
\begin{figure}
\centering
\includegraphics[width=0.45\textwidth]{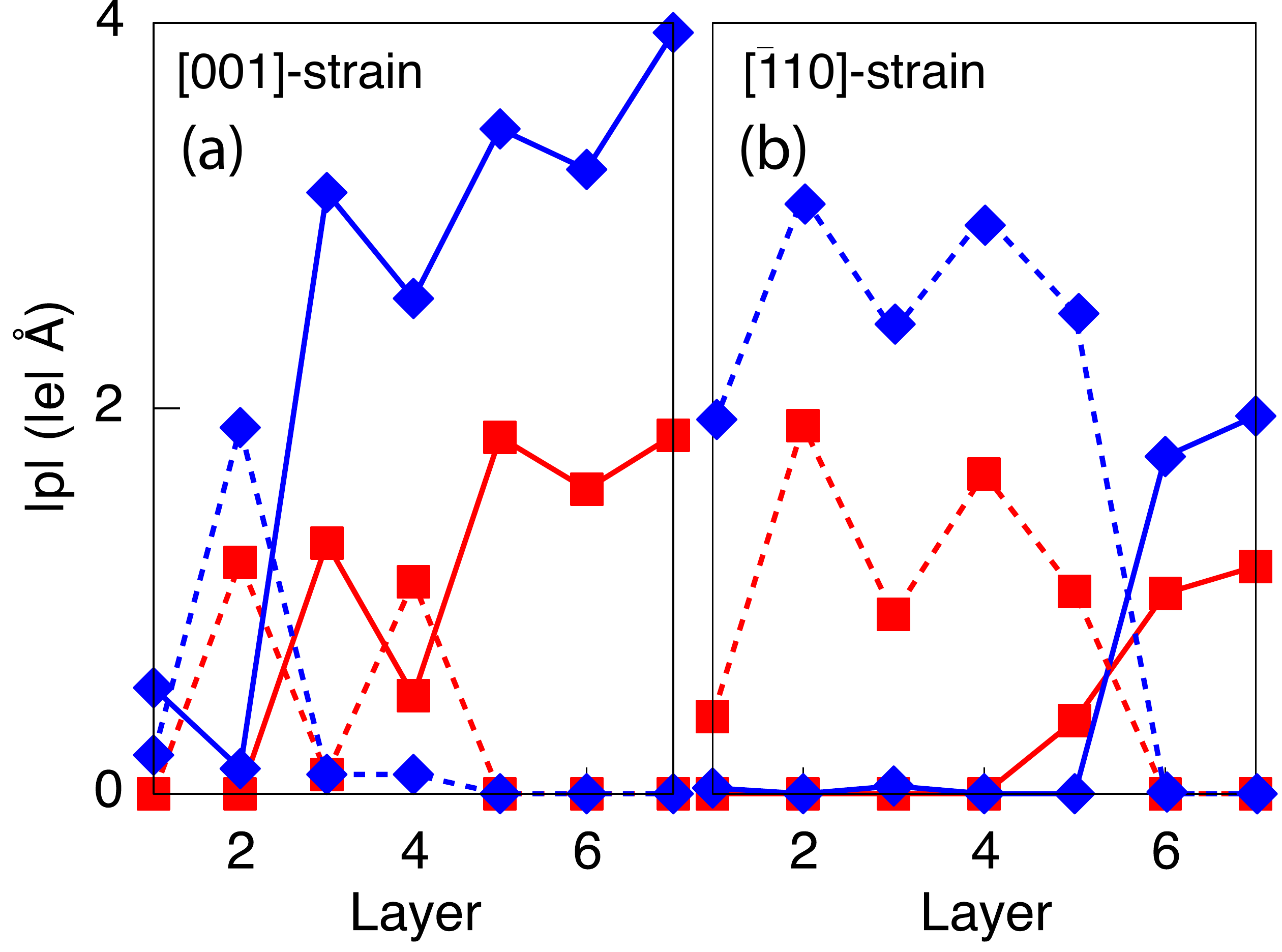}
\caption{(Color online) Layer-resolved dipole moments under tensile
  uniaxial strain for an optimization of the atomic structure without
  imposed symmetry. The substrate layers have been fixed to the
  corresponding polar states along (001).  Solid lines: $|p_{[001]}|$,
  dashed lines: $|p_{[\bar{1}10]}|$. 2~\% strain (red, squares); 5~\%
  strain (blue, diamonds); (a) [001] strain; (b) $[\bar{1}10]$
  strain.}
\label{fig:dipolesym}
\end{figure}

In the following, we present our results on polar atomic distortions
of the clamped surface.  For the unstrained film we tested the
stability of polar distortions along [001] by a displacement of
0.1~{\AA} along the polar A$_{2u}$ bulk mode of the 2 clamped bottom layers, {\it{cf.}}
Sec.~\ref{subsec:modes}.  As the upper layers are fully
relaxed, this setup corresponds to the clamping of the TiO$_2$ film to
a ferroelectric substrate.  Although, small polar displacements are
stable with respect to the paraelectric configuration in layer 3-5,
the amplitude of these polar displacements decays with the distance
towards the polar layers and thus no stable polarization is obtained
directly at the surface, see Fig.~\ref{fig:dipole}(a).  This decline
of the polarization shows that the ferroelectric state is not stable
within the unstrained film although a large polarizability is present.
Furthermore, the decay of the polar distortions is even more
pronounced if compressive [001] strain is imposed.  This indicates
that a ferroelectric distortion might be stabilized by a systematic
expansion of the Ti-O distances, {\it{cf.}} discussion of the bulk
case in Sec~\ref{subsec:modes}.

Indeed, polar distortions along [001] or [$\bar{1}$10] are stable
within the TiO$_2$ surface for 2~\% uniaxial tensile strain, see
Fig.~\ref{fig:dipole}.  Additionally, the short-range repulsion is
further reduced for an increasing tensile strain and thus the local
dipole moments increase systematically if we impose an uniaxial
expansion of 5~\%.

Table~\ref{tab:E} lists the energy gain for the different polar
displacements in comparison to the strained paraelectric films.  In
agreement with the bulk system, both polar states along [001] and
[$\bar{1}$10] are energetically more favorable than the paraelectric
state and thus two stable ferroelectric states exist.  Obviously, the
polar distortions parallel to the imposed strain are energetically
most stable. But, in agreement to the strained bulk, the paraelectric
state is also unstable against a polar distortion in the direction
perpendicular to the strain, see Tab.~\ref{tab:umod}.  Only for 2~\%
[001] strain the paraelectric state is more favorable than the
ferroelectric state perpendicular to the strain direction.

For the obtained dipole patterns in Fig.~\ref{fig:dipole}, we
constrained the atomic optimization and enforced the local dipoles to
point parallel to the fixed bottom layer.  In contrast, if all atomic
degrees of freedom are optimized without such constraint, local
dipoles both along and perpendicular to the polarization of the fixed
bottom layer form, see Fig.~\ref{fig:dipolesym}. For the polar states
along [001], the system obviously gains energy, if superimposed polar
distortions along [$\bar{1}10$] are formed, Tab.\ref{tab:E}. The energies for polar distortions in both directions are given in brackets. The modification of the local dipoles is even more
pronounced under tensile strain along [$\bar{1}10$]. Polar distortions
along [001] can be stabilized under such strain if the symmetry of the
system is fixed, which could possibly be achieved by an external
electric field. Despite this, polar distortions parallel to the strain
direction are much more favorable and are even stable on top of a
ferroelectric substrate polarized along [001], see
Fig.~\ref{fig:dipolesym}(b).

\begin{table}
\caption{Energy gain relative to the corresponding paraelectric case
  (meV/2 f.u. surface cell) of the different polar surface
  configurations, see text. The bottom layers are fixed to the polar
  state along [001] ($[\bar{1}10]$) in the first (second) row.  Only
  in-plane displacements parallel to the bottom polarization are
  allowed. (The upper layers are fully relaxed resulting in local
  dipole moments along [001] and $[\bar{1}10]$).}
\begin{tabular}{lccccc}
\hline \hline
&\multicolumn{2}{c}{[001]-Strain}&\multicolumn{2}{c}{$[\bar{1}10]$-Strain}\\
 Bottom layers &2~\%&5~\%&2~\%&5~\%\\
 $[001]$ polarized&18 (25)&262 (275)&3 (48)&111 (518)\\
$[\bar{1}10]$ polarized&-1~~~~~&~~48 (128)&28~~~~~~~~&525~~~~~~~~\\
 \hline
\hline
\end{tabular}
\label{tab:E}
\end{table}

We now discuss the observed layer-dependent variations of the local
dipoles. In all cases, the local dipole moments in [001]
($[\bar{1}10]$) direction are enlarged in layers with an odd (even)
number. Besides this, the local polarization in [001] direction
slightly decreases towards the surface and is quenched within the
first first layer for $[\bar{1}10]$ strain or 2~\% expansive [001]
strain. For the local polarization along $[\bar{1}10]$ no systematic
decay of the polarization towards the surface is obtained, although,
the polarization in the first layer is considerably reduced for all
kinds of strain.

The different Ti-O distances and thus the local variations of the
short-range repulsion have a large influence on the obtained dipole
patterns, see discussions in Sec.~\ref{subsec:modes} and  Sec.~\ref{subsec:free}.  The Ti-O distances at the surface and thus
the short-range repulsion oscillates with the distance form the
surface.  Additionally, the tensile strain further modifies the
spatial variation of the Ti-O distances.  First of all, the
Ti-O$_{eq}$ distances within the Ti-O$_p$ planes increase
systematically with both kinds of tensile strain, see
Tab.~\ref{tab:strain}. In addition, the apical distances in-plane
increase with [$\bar{1}$10] strain whilst they are not modified by
strain along [001].  Because of this, the short-range repulsion which
builds up during a polar shift between Ti atoms and their O$_p$
neighbors in [001] and $[\bar{1}10]$ direction is reduced considerably
and thus polar distortions are stabilized by tensile strain.

As the tensile in-plane strain enlarges the atomic volume, a shrink of
the atomic distances along the surface normal is very likely. Indeed,
the Ti-Ti interlayer distances decrease for tensile [001] strain.
 In contrast to this, no
unique reduction of the interlayer distances can be found for tensile
[$\bar{1}$10] strain.

Additionally, for the Ti-O bonds oriented perpendicular to the
surface, the discussed strengthening (weakening) of the bond has a
significant impact on the atomic relaxation under imposed strain. On
one hand, the strengthened bonds are rather insensitive to tensile
strain. In particular, the distance between the Ti$_{6c}$ atom in
layer 1 and the outer O$_b$ atom is not modified by 2~\% [001] strain
or [$\bar{1}$10] strain and shows an expansion of only 1~\% under 5~\%
[001] strain.  As a consequence, a polar displacement of this
Ti$_{6c}$ atom relative to the surrounding atoms along [001] is not
stable, even at 5~\% expansive strain in either direction.  On the
other hand, the length of the less covalent bonds increases
(decreases) systematically under expansive [001] ([$\bar{1}$10])
strain.  Because of this, the asymmetry of the Ti-O bond lengths along
the surface normal increases (decreases) slightly with the imposed
[001] ([$\bar{1}$10]) strain for each Ti atom.

Besides the modified short-range repulsion, the dynamic charges at the
rutile surface oscillate with the distance from the surface, see
Fig.~\ref{fig:bornlayer}.  Consequently, the local dipole moments show
the same trends with maximal (minimal) local dipole moments along
[001] in odd (even) layers and the other way around for polarization
along [$\bar{1}$10].  Here, the magnitude of Z$^*$ has a direct
influence on the local dipole moments, see Eq.~\ref{eq:dipol}. Also,
the local variation of the dipolar interaction leads to a modulation
of the atomic displacements.\\
Thus, the considerable difference between Z$^*_{||}$ and Z$^*_{\perp}$
leads to different amplitudes of the [$\bar{1}$10] displacements of
Ti$_{5c}$ and Ti$_{6c}$ atoms in each layer. For example, the apical
bond of the Ti$_{6c}$ atoms in the first layer is aligned along
[$\bar{1}$10] and thus the displacement of this atom against the
surrounding oxygen octahedra is a factor of two larger than those of
the Ti$_{5c}$ atom within the same layer. Analogously, the
displacement of the Ti$_{5c}$ atom in each even layer is a factor of
two larger than the displacements of the corresponding Ti$_{6c}$ atom.
Additionally, Z$_{||}$ is considerably larger within the even layers,
as are the amplitudes of the polar distortions. Similarly, the
reduction of $\vec{p}_{[001]}$ in the layers with an even number is
due to a reduced displacement of the Ti$_{6c}$ atoms within these
layers, which is correlated with the reduced dipolar interaction along
[001] for these atoms.

Both Z$^*_{[001]}$ and Z$^*_{[\bar{1}10]}$ are slightly modified under
tensile strain, see Tab.~\ref{tab:bornstrain}. For example,
Z$^*_{[001]}$ of the O$_b$ atoms increases (decreases) for the
strengthened (weakened) bonds, while the Born charge of the O$_p$
atoms is insensitive towards such strain. As a consequence, the
oscillation of the dipolar interaction is slightly enlarged in
comparison to the equilibrium lattice constant. For [$\bar{1}$10]
strain Z$^*_{[001]}$ of all O$_p$ and Ti atoms is enlarged and thus
the magnitude of the Z$^*_{[001]}$ oscillation is reduced.
Simultaneously, the layer-wise oscillation of p$_{[001]}$ is smaller
for [$\bar{1}$10] strain.

\begin{table*}
\caption{Born charges ($|e|$) within the paraelectric phase of the
  rutile (110) surface calculated with VASP for different values of
  uniaxial [001] and [$\bar{1}10$] strain.  For the Ti$_{6c}$/Ti$_{5c}$ atoms and the Ti atoms below, respectively the O$_p$ and
  O$_b$ atoms, the Born charge for the first free surface layers are
  given, see Fig.~\ref{fig:obstruk}.  Upper part Z$^*_{[001]}$; Lower
  part: Z$^*_{[\bar{1}10]}$.  }

\begin{tabular}{ccccccccccccc}
\hline
\hline
&\multicolumn{3}{c}{Ti$_{6c}$}&\multicolumn{3}{c}{Ti$_{5c}$}&\multicolumn{3}{c}{O$_{p}$}&\multicolumn{3}{c}{O$_{b}$}\\
Strain&1&2&3&1&2&3&1&2&3&1&2&3\\
\hline
No strain&~~8.4~~&~~7.3~~&~~8.3~~&~~7.3~~&~~7.6~~&~~8.2~~&~~-3.8~~&~~-3.7~~&~~-4.1~~&-5.6/-2.8&-2.8/-4.5&-4.6/-3.7\\
$[001]_{+2\%}$&~~8.6~~&~~7.2~~&~~8.3~~&~~7.3~~&~~7.6~~&~~8.2~~&~~-3.8~~&~~-3.7~~&~~-4.1~~&-5.8/-2.7&-2.7/-4.6&-4.7/-3.6\\
$[001]_{+5\%}$&~~8.7~~&~~7.1~~&~~8.4~~&~~7.2~~&~~7.6~~&~~8.1~~&~~-3.8~~&~~-3.6~~&~~-4.1~~&-6.1/-2.6&-2.5/-4.7&-4.8/-3.5\\
$[\bar{1}10]_{+2\%}$&~~8.6~~&~~7.3~~&~~8.4~~&~~7.4~~&~~7.7~~&~~8.3~~&~~-3.8~~&~~-3.7~~&~~-4.2~~&-5.7/-2.9&-2.9/-4.5&-4.7/-3.8\\
$[\bar{1}10]_{+5\%}$&~~8.9~~&~~7.5~~&~~8.7~~&~~7.6~~&~~8.0~~&~~8.4~~&~~-3.9~~&~~-3.8~~&~~-4.2~~&-5.7/-3.1&-3.1/-4.6&-4.7/-4.0\\
\hline
No strain&~~6.3~~&~~5.7~~&~~7.2~~&~~4.5~~&~~7.8~~&~~5.1~~&~~-4.3~~&~~-5.5~~&~~-4.9~~&-1.2/-1.3&-1.1/-1.2&-1.4/-1.4\\
$[001]_{+2\%}$&~~6.4~~&~~5.7~~&~~7.4~~&~~4.6~~&~~7.9~~&~~5.1~~&~~-4.4~~&~~-5.6~~&~~-5.0~~&-1.1/-1.3&-1.1/-1.2&-1.4/-1.4\\
$[001]_{+5\%}$&~~6.7~~&~~5.9~~&~~7.6~~&~~4.7~~&~~8.0~~&~~5.3~~&~~-4.6~~&~~-5.6~~&~~-5.2~~&-1.2/-1.3&-1.1/-1.2&-1.4/-1.4\\
$[\bar{1}10]$$_{+2\%}$&~~6.1~~&~~5.7~~&~~7.0~~&~~4.4~~&~~7.8~~&~~4.9~~&-4.1&-5.5&-4.8&-1.1/-1.3&-1.1/-1.2&-1.4/-1.4\\
$[\bar{1}10$]$_{+5\%}$&~~5.9~~&~~5.7~~&~~6.7~~&~~4.2~~&~~7.6~~&~~4.8~~&~~-4.0~~&~~-5.3~~&~~-4.6~~&-1.2/-1.3&-1.1/-1.2&-1.4/-1.3\\
\hline
\hline
\end{tabular}
\label{tab:bornstrain}
\end{table*}

\section{Conclusions and Outlook}
\label{sec:outlook}
We have investigated the ferroelectric trends of the incipient
ferroelectric material TiO$_2$ rutile by means of first-principles
calculations. Similar to the well-known ferroelectric perovskites,
rutile possesses anomalously large dynamic charges, which is
one key factor for the stabilization of a ferroelectric phase.

We could show that this strong dynamic charge transfer is mainly
mediated by $\pi$-type O-Ti hybrid orbitals. Furthermore, this dynamic
charge transfer is considerably modified by changing the ratio of the
equatorial and apical Ti-O bond lengths. Nevertheless, a large dynamic
charge transfer along the [001] direction and along the apical Ti-O
bonds is preserved under small variations of the lattice as they may
appear at strained surfaces. Indeed, the Born charges in both
directions are quite stable at the (110) surface of rutile, even if
one imposes an uniaxial tensile strain in the surface planes. Only
small oscillations of the dipolar interaction appear with the surface
distance. Here, the dipolar interaction along [001] is slightly
enhanced within every second Ti-O layer starting from the topmost
layer, while the dipolar interaction along $[\bar{1}10]$ is slightly
larger in the odd layers.

Although, the material possesses such a large dynamic charge transfer,
no ferroelectric transition appears, as the short-range repulsion,
which builds up during a polar displacement is too large for the
undistorted material. In agreement with previous theoretical studies,
we have shown that a ferroelectric transition of the bulk material can
be enforced by lattice expansion or by uniaxial tensile strain. In
both cases, the enlarged Ti-O distances reduce the short-range
repulsion during a polar displacement.  Additionally, we could show
that a ferroelectric state polarized along [001] can be stabilized by
all modifications of the lattice which enlarge the nearest Ti-O
distance, even if the apical Ti-O distance shrinks during the lattice
modification.  For uniaxial strain along [001] and [$\bar{1}10]$
stable ferroelectric states can be found and for both kinds of strain,
the polar displacement along the strain direction is most favorable.

Most notably, the stabilization of two ferroelectric states, one
polarized along [001] and one along [$\bar{1}10]$ can be transferred
from the bulk material to the rutile (110) surface.  We thus find
large local dipole moments already for 2~\% tensile strain in both
[001] and [$\bar{1}10$] direction. Although the ferroelectric trends
and the obtained local dipole moments may be slightly overestimated
within our theoretical approach, we find a successive increase of the
local dipole moments for increasing strain. Therefore, we predict that
a ferroelectric state can be stabilized for experimentally accessible
tensile strains, and that the resulting electric dipoles persist even
in the atomic layers at or immediately below the surface. On the other
hand, for compressive strain no stabilization of a ferroelectric phase
is possible at the rutile surface, in agreement with the corresponding
bulk material.

\section{Acknowledgments}
A. Gr{\"u}nebohm and P. Entel acknowledge financial support by the
Deutsche Forschungsgemeinschaft (SFB 445) and C. Ederer by Science
Foundation Ireland (SFI-07/YI1/I1051). The authors thank the John von
Neumann Institute of Computing (NIC) and the J\"ulich Supercomputing
Center (JSC), as well as the Center of Computational Science and
Simulation (CCSS) of the University of Duisburg-Essen for computing
time. Some calculations have also been performed at the Trinity Centre
for High Performance Computing (TCHPC).


\begin{thebibliography}{43}
\bibitem[{hil()}]{hilfe}
\bibinfo{note}{We note that an optimization of the lattice vector in [110]
  direction under 2~\% expansion in [$\bar{1}$10]/[001] direction results in a
  decrease of only 0.2~\%/0.4~\% for the paraelectric state.}
\end{thebibliography}
\end{document}